\documentclass[nofootinbib,aps,pre,twocolumn,secnumarabic,balancelastpage,amsmath,amssymb,floatfix,superscriptaddress,10pt]{revtex4-2}

\usepackage{xcolor}

\usepackage{bm}
\usepackage{mathtools}

\newcommand{\tr}{\operatorname{tr}}
\newcommand{\ketbra}[2]{\ket{#1} \hskip -0.8ex \bra{#2}}

\newcommand{\up}{\uparrow}
\newcommand{\dn}{\downarrow}
\newcommand{\expct}[1]{\left\langle #1 \right\rangle}

\usepackage{amssymb}
\usepackage{amsmath}
\usepackage{amsfonts}
\usepackage{braket}
\usepackage{graphicx}
\usepackage{epstopdf}
\usepackage{epsfig}
\usepackage{dsfont,amsthm,amsbsy}
\usepackage{hyperref}
\usepackage[xspace]{ellipsis}
\usepackage{cancel}
\usepackage[version=4]{mhchem}
\usepackage{algorithm}
\usepackage{algpseudocode}

\DeclareMathOperator{\argmax}{argmax}

\DeclareMathOperator{\erf}{erf}

\newcommand{\av}{\text{av}}
\newcommand{\bulk}{\text{bulk}}

\newcommand{\antigs}{\text{anti-g.s.}}
\newcommand{\gs}{\text{g.s.}}
\newcommand{\para}{\text{para}}
\newcommand{\Heis}{\text{Heis.}}
\newcommand{\TFIM}{\text{TFIM}}
\newcommand{\live}{\text{live}}
\newcommand{\nlive}{n_\text{live}}
\newcommand{\Haar}{\text{Haar}}
\newcommand{\haarint}[1]{\int_{\Haar}\hskip -2.5ex d{#1}\;}
\newcommand{\gmc}{Galilean Monte Carlo}
\newcommand{\GbE}{G$\beta$E}

\newcommand{\bin}{\text{bin}}

\makeatletter
\newcommand{\appendixsubsectionformat}{%
  \begingroup
  \renewcommand{\p@subsection}{\thesection.}
  \renewcommand{\p@subsubsection}{\thesection.\thesubsection.}
}
\makeatother

\definecolor{dkred}{rgb}{0.5,0,0}
\definecolor{dkgreen}{rgb}{0,0.5,0}

\begin{document}

\title{Eigenstate condensation in quantum systems with finite-dimensional Hilbert spaces}

\author{Christopher David White}
\email{christopher.d.white117.ctr@us.navy.mil}
\affiliation{
    Center for Materials Physics and Technology, U. S. Naval Research Laboratory, Washington, DC 20375, USA
}

\author{Michael Winer}
\affiliation{Institute for Advanced Study, Princeton, NJ 08540, USA }
\affiliation{Alignment Research Center, Berkeley, CA 94704}

\author{Noam Bernstein}
\affiliation{
    Center for Materials Physics and Technology, U. S. Naval Research Laboratory, Washington, DC 20375, USA
}

\begin{abstract}
Random quantum states drawn from the Haar ensemble with a constraint on the energy expectation value $E_\av = \braket{\psi | H | \psi}$ display \textit{eigenstate condensation}:
for $E_\av$ below a critical value $E_{c-}$ or above $E_{c+}$,
they develop macroscopic overlap with the ground state or anti-ground-state.
We use analytical calculations and state-of-the-art numerical methods to investigate the eigenstate condensation phase transition.
We give compact expressions for the critical energies;
derive an analytical scaling form for the order parameter
in systems with an extensive free energy;
show that in those systems, the phase transition itself has exponential rather than power-law finite size scaling;
and test these results with large-scale numerical sampling in random matrices and a range of local spin Hamiltonians.
In local spin systems the two critical energies approach the middle of the spectrum as $1/V$
with $V$ the number of spins.
Our scaling form, however, demonstrates that the condensation phase transitions have exponential, rather than polynomial, finite-size scaling,
which justifies treating the high-temperature phase as an extended phase.
\end{abstract}

\maketitle

\begin{figure*}[t]
    \centering
    \includegraphics[width=0.98\textwidth]{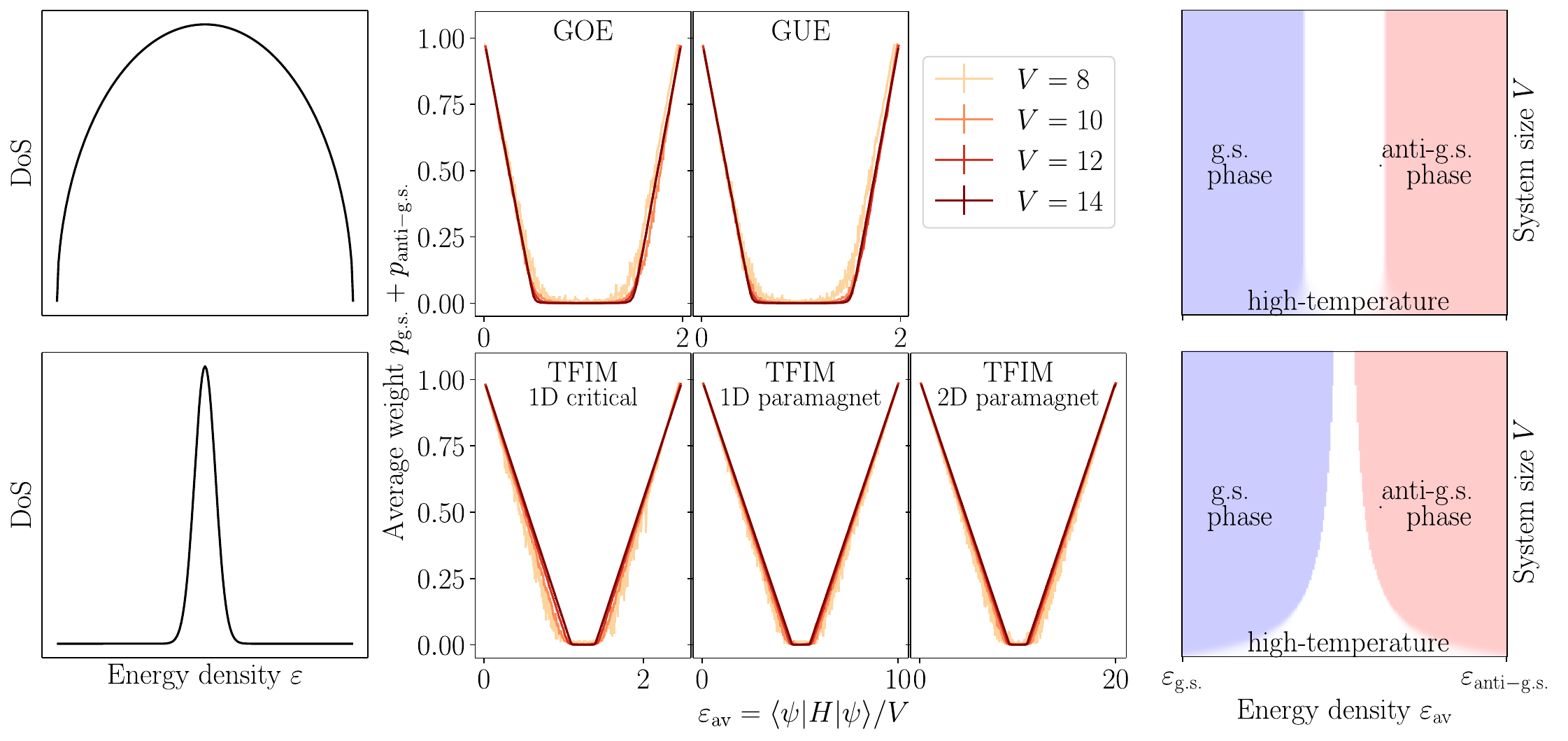}
    \caption{
    \textbf{Eigenstate condensation} in five models: GOE and GUE random matrix Hamiltonians; the paramagnetic and critical transverse field Ising model (TFIM) in 1D, and the paramagnetic TFIM in 2D.
    \textbf{Left column}: (heuristic) density of states.
    \textbf{Center:} combined ground state and anti-ground-state probability $p_\gs + p_\antigs = |\braket{\gs | \psi}|^2 + |\braket{\antigs|\psi}|^2$ across system size;
    in each case we see $O(1)$ weight on the ground state for energy densities $\varepsilon_\av < \varepsilon_{c-}$ and on the anti-ground-state for $\varepsilon_\av > \varepsilon_{c+}$.
    \textbf{Right:} (heuristic) phase diagrams.
    \textbf{Top row:} random matrices have a semicircle density of states;
    consequently the critical energy densities approach $\varepsilon_{c-} = 1/2, \varepsilon_{c+} = 3/2$ for large systems.
    \textbf{Bottom row:} local spin systems have Gaussian densities of states with energy-density variance shrinking with system size $V$ like $1/V$.
    Consequently the ground-state and anti-ground-state phases approach the middle of the spectrum.
    }
    \label{fig:fig1}
\end{figure*}

\section{Introduction}

Random quantum states are ubiquitous in the study of thermodynamics and thermalization.
In quantum statistical mechanics, microcanonical averages are averages over random states in the span of the energy eigenstates in a narrow energy window
\cite{
schrodingerStatisticalThermodynamicsCourse1952,%
khinchinMathematicalFoundationsQuantum1960%
}.
Indeed it has been rigorously shown that a state chosen from the Haar distribution over the vector space spanned by states in a small energy window approximates this microcanonical average, and therefore the Gibbs state on a subsystem
\cite{bocchieriErgodicFoundationQuantum1959a,%
lloydPureStateQuantum2013,%
goldsteinCanonicalTypicality2006a,%
lloydExcuseOurIgnorance2006,%
popescuEntanglementFoundationsStatistical2006};
this is called ``canonical typicality''.
Thermalization of an initial state under time evolution is also understood in terms of the eigenstate thermalization hypothesis
\cite{deutschQuantumStatisticalMechanics1991,srednickiChaosQuantumThermalization1994,rigolThermalizationItsMechanism2008,dalessioQuantumChaosEigenstate2016,deutschEigenstateThermalizationHypothesis2018,uedaQuantumEquilibrationThermalization2020}
and random states subject to constraints on energy eigenstate overlaps
\cite{markMaximumEntropyPrinciple2024}
or (heuristically) the first few moments of the energy 
\cite{dalessioQuantumChaosEigenstate2016}.

But a quantum state drawn at random subject only to a constraint on its energy expectation value $E_\av = \braket{\psi | H | \psi}$
is far from thermal
\cite{woottersRandomQuantumStates1990,brodyQuantumCanonicalEnsemble1998,brodyMicrocanonicalDistributionsQuantum2005,benderSolvableModelQuantum2005a,naudtsGeneralizedQuantumMicrocanonical2006a,jona-lasinioStatisticsQuantumExpectations2006,fineTypicalStateIsolated2009,brodyQuantumMicrocanonicalEquilibrium2007,brodyQuantumPhaseTransitions2007,fineAlternativeConventionalMicrocanonical2012a,freschTypicalityEnsemblesQuantum2009,freschEmergenceEquilibriumThermodynamic2010,freschEmergenceEquilibriumThermodynamic2010a,freschQuantumMicrocanonicalStatistics2011,jiNonthermalStatisticsIsolated2011,campisiQuantumFluctuationRelations2013,alonsoNonextensiveThermodynamicFunctions2015,reimannWhyAreMacroscopic2020,millerStatisticalMechanicsRandom2025}.
It is broadly distributed over energy eigenstates%
---$p_\alpha \equiv |\braket{E_\alpha|\psi}|^2 \sim 1 / (1 + \beta E_\alpha)$ for a temperature-like parameter $\beta$---%
so the single-state energy uncertainty is large \cite{woottersRandomQuantumStates1990,fineTypicalStateIsolated2009}.
And for $E_\av$ below some critical $E_c$ it undergoes a phase transition \cite{fineTypicalStateIsolated2009,reimannFullExpectationvalueStatistics2019} that we call \textit{eigenstate condensation}.
Eigenstate condensation is structurally similar to Bose-Einstein condensation;
in it energy-constrained random states develop macroscopic overlap with the ground state or anti-ground-state.
In finite Hilbert spaces, then, the energy-constrained ensemble generically displays three phases (Fig.~\ref{fig:fig1}):
a ground-state-like phase;
a high-temperature phase, in which they have exponentially small overlap with each energy eigenstate;
and an anti-ground-state phase.
When the Hamiltonian comes from the Gaussian orthogonal ensemble,
this transition maps to the replica transition found by Kosterlitz, Thouless, and Jones
for the classical \(p = 2\) $p$-spherical spin model \cite{kosterlitzSphericalModelSpinGlass1976}.

The eigenstate condensation transition is difficult to study numerically because  energy-constrained random states are difficult to sample \cite{hantschelMonteCarloSampling2011}.
For spin systems, Haar-random states with energy expectation value away from mid-spectrum are doubly exponentially rare:
unconstrained Haar-random states on a finite-dimensional Hilbert space have an approximately Gaussian distribution of energies
\cite{auerbachInteractingElectronsQuantum1998,camposvenutiProbabilityDensityQuantum2013,reimannFullExpectationvalueStatistics2019},
and the width of the Gaussian is itself inversely proportional to Hilbert space size and so exponentially small in system size.
Prior work has therefore been limited to Hilbert space dimensions $\sim O(10)$, i.e. very small spin systems.

In this work we identify the critical condensation energies;
derive an analytical scaling form for the order parameter
in systems with an extensive free energy;
show that in those systems, eigenstate condensation has exponential rather than power-law finite size scaling;
and test these results with large-scale numerical sampling in random matrices and a range of local spin Hamiltonians.
We do so by combining 
analytical calculations with two numerical techniques originating in Bayesian statistics,
nested sampling
\cite{skillingNestedSampling2004,skillingNestedSamplingGeneral2006,ashtonNestedSamplingPhysical2022}
and Galilean Monte Carlo
\cite{skillingBayesianComputationBig2012,betancourtNestedSamplingConstrained2011a}
applied to systems of Hilbert space dimension up to $2^{14} = 16384$.
For Hamiltonians from the Gaussian orthogonal or unitary ensemble (GOE or GUE),
the two critical energy densities, separating the high-temperature phase from the ground state and anti-ground-state phases, approach a constant. 
For systems with extensive free energy, by contrast---this includes many local spin systems, like the Ising and Heisenberg models---%
the two critical energies approach the middle of the spectrum as $1/V$, $V$ the number of spins.
For these systems, then, in the large-system limit the high-temperature phase is not an extended phase but a point:
at almost any energy density, a random state has large overlap with either the ground state or the most highly excited state.
Nonetheless the high-temperature phase is best understood as an extended phase:
our scaling form demonstrates that the critical region around the transition shrinks not polynomially but exponentially, like $1 / \sqrt {V 2^{V}}$.
These phenomena occur across systems: in 
random matrices;
free-fermion Hamiltonians;
and spin systems that are 
one dimensional and two dimensional,
gapped and gapless,
and non-integrable and Bethe ansatz integrable.

\section{Models}
We work with random matrices and spin Hamiltonians on finite Hilbert spaces of dimension $\mathcal N = 2^V$.
Our analytical results depend only on the density of states, so they apply to
any spin system with extensive high-temperature free energy.

In numerics we consider random GOE and GUE Hamiltonians:
\begin{align}\label{eq:gbe}
    H = \frac {V} {2^{3/2}\sqrt{\mathcal N}}\sum_{\alpha,\beta = 1}^{\mathcal N} \left(X_{\alpha\beta} + \overline{X_{\beta \alpha}}\right) \ketbra{\alpha}{\beta} + E_0
\end{align}
with $X_{\alpha\beta}$ i.i.d. from the standard real or complex Gaussian distribution, for GOE or GUE respectively.
For large $V$, hence large $\mathcal N$, the density of states is given by the famous Wigner semicircle law \cite{mehta2004,tao2012}.
We choose the scaling in \eqref{eq:gbe} so the semicircle has diameter exactly $2V$;
we shift (choose $E_0$) so the ground state has $E_\gs = 0$.

We also consider a family of spin systems, the transverse-field Ising model (TFIM) 
\begin{align}\label{eq:ham}
    H_{\TFIM} &= \sum_{\langle ij\rangle} \sigma^z_i \sigma^z_{j} +\sum_j h_x \sigma^x_j + E_0
\end{align}
on square lattices of dimension $D = 1,2$.%
\footnote{In App.~\ref{app:add-ham} we additionally discuss the 1D mixed-field Ising model as well as the 1D Heisenberg model, which is Bethe ansatz integrable.
The Heisenberg model displays less favorable finite-size scaling, because its density of states does not approach the true large-system Gaussian for accessible system sizes.}
In one dimension the TFIM is free-fermion integrable and has a quantum critical point  at $h_x = 1$;
in two dimensions it is non-integrable and satisfies the eigenstate thermalization hypothesis.
For $h_x \ll 1$ it has an exponentially small gap between nearly-degenerate cat states $\ket{\up \dots \up} \pm \ket{\dn \dots \dn}$.

In 1D we consider the paramagnetic $h_x = 5$, critical $h_x = 1$, and ferromagnetic $h_x = 0.2$ TFIM.
In 2D, where the TFIM has a zero-temperature phase transition at $h_x \approx 3.04$ \cite{pfeutyIsingModelTransverse1971,jonghCriticalBehaviour2D1998,riegerApplicationContinuousTime1999}, we take $h_x = {10}$;
we do not investigate the interplay of the eigenstate condensation phase transition and the finite-temperature ferromagnetic phase transition.
In each case we choose the constant shift $E_0$ so that the ground state energy $E_{g.s.} = 0$.

\section{Numerical methods}
We use nested sampling
\cite{skillingNestedSampling2004,skillingNestedSamplingGeneral2006,ashtonNestedSamplingPhysical2022}
with Galilean Monte Carlo
\cite{betancourtNestedSamplingConstrained2011a,skillingBayesianComputationBig2012}
to treat systems up to $V = {14}$ spins, with Hilbert spaces up to dimension $\mathcal N = {16384}$.
Nested sampling iteratively modifies a pool of states. 
At each iteration it removes and records the state with the largest energy expectation value, denoted $E_* = \max_{\psi \in \text{pool}} \braket{\psi | H | \psi}$,
and draws a new state $\ket {\psi'}$ (``resamples'') from the constrained manifold of states with energy expectation value $\braket{\psi' | H | \psi'} < E_*$.

The efficiency of nested sampling depends on the efficiency of the resampling procedure;
we use Galilean Monte Carlo (GMC) \cite{skillingBayesianComputationBig2012}.
In Galilean Monte Carlo a single state from the pool is moved in a random direction through a constrained manifold (in our case, the manifold of states $\ket{\psi'}$ below the energy cutoff $\braket{\psi' | H|\psi'} < E_*$),
specularly reflecting from the boundaries given by the constraint.
The state therefore ``advects'' through the constrained manifold,
efficiently sampling it.
(We describe nested sampling and Galilean Monte Carlo in more detail in the appendices.)

\begin{figure}[t]
    \centering
    \includegraphics[width=0.95\columnwidth]{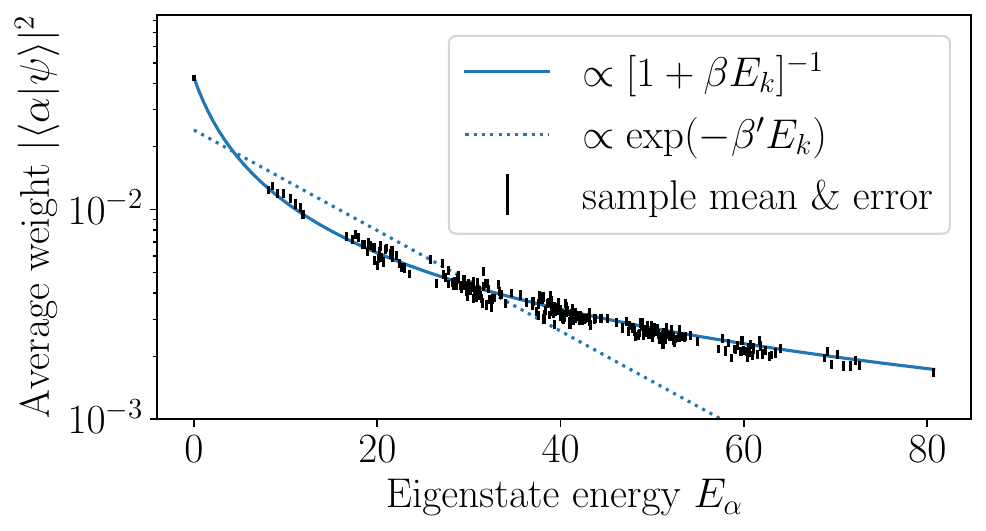}
    \caption{\textbf{Average weight on energy eigenstates} for $400$ random states with energy $E_{\av} \approx {33}$ in the $V = {8}$ paramagnetic TFIM.
    The weight is well-described by $p_k \propto [1 + \beta E_k]^{-1}$, but not by the Boltzmann probability $p_k \propto e^{-\beta' E_k}$ (best fit $\beta, \beta'$ shown in blue).
    }
    \label{fig:energy-eigenstate-weight}
\end{figure}

Nested sampling with GMC confirms the sharply non-Gibbs eigenstate weights of energy-constrained random states.
Those weights are (on average) \cite{woottersRandomQuantumStates1990,fineTypicalStateIsolated2009}
\begin{align}
\label{eq:pk-Ek}
\begin{split}
    p_\alpha \equiv |\braket{E_\alpha|\psi}|^2 &= Z^{-1} \frac 1 {1 +\beta E_\alpha}\\
    Z(\beta) &= \sum_\alpha \frac 1 {1 + \beta E_\alpha}\;,
\end{split}
\end{align}
where the parameter $\beta$ depends on the constraint energy $E_\av = \braket{\psi | H |\psi}$
and the normalization function $Z(\beta)$ is closely related to the resolvent of the Hamiltonian.
(This is in sharp contrast to a Gibbs weight $p_\alpha \propto e^{-\beta E_\alpha}$.)
We give a compact derivation of this form in App.~\ref{app:ed}.
Fig.~\ref{fig:energy-eigenstate-weight} shows $p_\alpha$ for the paramagnetic TFIM; the average sample weights are consistent with \eqref{eq:pk-Ek}.

\section{Condensation on ground state or anti-ground-state}
As the energy expectation value $E_\av$ decreases, the system suddenly develops $O(1)$ weight on the ground state \cite{fineTypicalStateIsolated2009}.
As \cite{reimannFullExpectationvalueStatistics2019} observed, this is in fact a phase transition:
an order parameter, the ground state weight $p_\gs$, is non-analytic in the state energy expectation value.%
The control parameter, the energy expectation value, controls the ensemble in precisely the same way that an ordinary microcanonical ensemble is controlled by the energy.
To understand how this comes about and compute the critical energy,
work first at fixed energy expectation value $E_\av < E_\infty = \mathcal N^{-1}\tr H$.
If the system has a non-degenerate ground state $\ket 0$ with a gap $E_1 - E_0 > O(\mathcal N^{-1})$,
we can split its spectrum into ground state and a ``bulk'' consisting of the other $\mathcal N -1$ states.
The energy expectation value $E_\av$ and the inverse energy scale $\beta$ of a random state are related by
\begin{align} \label{eq:bulk-gs-av}
\begin{split}
    E_\av &= (1 - p_{\gs})  Z_{\bulk}(\beta)^{-1} \sum_{\alpha \ne \gs} \frac {E_\alpha} {1 + \beta E_\alpha} \\
    Z_{\bulk}(\beta) &=  \sum_{\alpha \ne \gs} \frac 1 {1 + \beta E_\alpha}%
\end{split}
\end{align}
If $p_\gs \sim \mathcal N^{-1}$ it can be neglected, and the energy is bounded below $ E_\av > \lim_{\beta \to \infty} \mathcal N[\beta Z(\beta)]^{-1}$; we label
\begin{align}\label{eq:Ec-sum}
\begin{split}
    E_{c-}^{-1} \equiv \lim_{\beta \to \infty} \frac {1}{\mathcal N} \beta Z_\bulk(\beta) &\approx \frac 1 {\mathcal N}\sum_{\alpha \ne \gs} \frac 1 {E_\alpha} %
\end{split}
\end{align}
(The bound on $E_\av$ is elementary but not immediate; see App.~\ref{app:Ebulk-bound} for details.)
Conversely if $E_\av < E_{c-}$ then $p_\gs$ must be $O(1)$,
and rearranging \eqref{eq:bulk-gs-av} in the large-$\beta$ limit gives
\begin{equation}\label{eq:pgs-pred}
    p_\gs = 1 - E_\av/E_c,\quad E_\av \ll E_c\;.
\end{equation}
$E_{c}$ depends only on the large-scale properties of the density of states $\rho(E)$---in particular the $E_c$ of Eq.~\eqref{eq:Ec-sum} is the first negative moment $\int dE\; \rho(E)/E$, excluding the ground state.
The same logic applies to the anti-ground-state: for $E_\av > E_{c+}$ defined analogously, the system develops macroscopic weight on the anti-ground-state.

Fig.~\ref{fig:fig1} shows $p_\gs + p_\antigs$ as a function of the energy density $\varepsilon_\av = E_\av / V$ for a variety of models and system sizes $V$.
Each model shows the tripartite structure predicted by \eqref{eq:pgs-pred} and its anti-ground-state analogue:
for $\varepsilon_\av < \varepsilon_{c-}$, $p_\gs + p_\antigs$ decreases linearly as $\varepsilon_\av$ increases;
for $\varepsilon_{c-} < \varepsilon_\av < \varepsilon_{c+}$, it is small; 
and for $\varepsilon_{c+} < \varepsilon_\av$ it increases linearly with $\varepsilon_\av$.

To understand how the critical energy depends on system size, use energy densities $\varepsilon = E/V$.
First consider GOE and GUE random matrices.
The Wigner semicircle density of states gives
\begin{align}\label{eq:GbE-Ec}
\varepsilon_{c-} = \frac 1 2\;, \quad \varepsilon_{c+} = \frac 3 2\;, 
\end{align}
in agreement with \cite{reimannFullExpectationvalueStatistics2019},
but for finite-size matrices the critical energy displays considerable variation
(cf App.~\ref{app:GbE}).

\begin{figure}[t]
    \centering
    \includegraphics[width=\columnwidth]{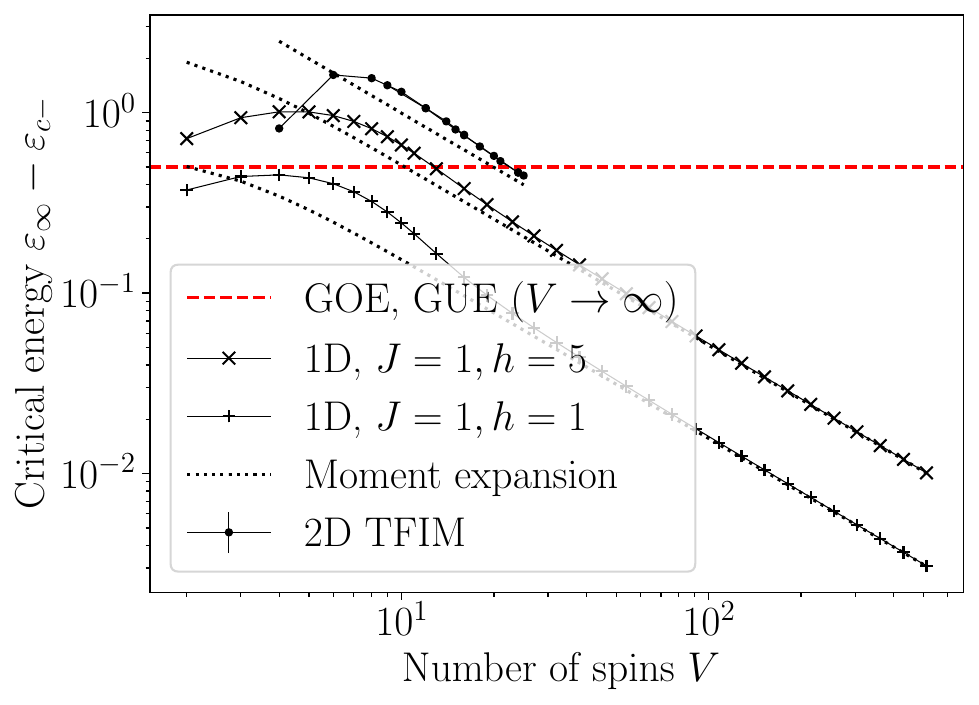}
    \caption{
    \textbf{Critical energy density} $\varepsilon_{c-}$ from \eqref{eq:Ec-sum}
    for {paramagnetic and critical} 1D Ising models and a paramagnetic 2D Ising model,
    together with the large-system limit for GOE and GUE matrices (Eq.~\ref{eq:GbE-Ec}).
    For the spin models a moment expansion (Eq.~\eqref{eq:Ec-large-L}; dotted lines) predicts that the critical energy density approaches the spectrum center $\varepsilon_\infty$ as $\varepsilon_\infty - \varepsilon_{c-} \propto 1/V$.
    For the Gaussian random matrix ensembles the Wigner semicircle predicts $\varepsilon_{c-} = 1/2$ independent of system size, though at finite size the random matrix $\varepsilon_{c\pm}$ exhibit considerable fluctuations (not shown).
    }
    
    \label{fig:Ec-L}
\end{figure}

For systems with extensive free energy---including many local spin Hamiltonians---the critical energy densities $\varepsilon_{c\pm}  = E_c / V$ approach the middle of the spectrum as $1/V$.
To see this, normalize  \eqref{eq:Ec-sum} by $V$ and expand the resulting $\varepsilon_\alpha$ around $\varepsilon_\infty \equiv (V\mathcal N)^{-1}\tr H $.
Because the free energy is extensive, the central moments of the density of states $\sum_\alpha (\varepsilon_\alpha - \varepsilon_\infty)^n$ fall off as $V^{-n/2}$ (App.~\ref{app:fe-mu}).
Including the leading-order correction, which comes from the second central moment $V s^2 = V \mathcal N^{-1}\sum_\alpha (\varepsilon_\alpha - \varepsilon_\infty)^2$, the critical energy density is
\begin{align}
\begin{split} \label{eq:Ec-large-L}
   \varepsilon_{c-} &\approx \varepsilon_\infty \left (1 - \frac {s^2}{V\varepsilon_\infty^2}\right) \\
   \varepsilon_{c+} &\approx \varepsilon_\infty \left (1 - \frac {s^2}{V(\varepsilon_\infty - \varepsilon_\antigs)^2}\right)
\end{split}
\end{align}
(cf App.~\ref{app:Ec-system-size}).
Fig.~\ref{fig:Ec-L} compares this prediction to exact calculations of \eqref{eq:Ec-sum} and calculations using Krylov methods;
we see good agreement for system sizes $V \gtrsim 10$.
(See App.~\ref{app:heisenberg} for the Heisenberg model, where subleading corrections complicate this picture.)

\begin{figure}
    \centering
    \includegraphics[width=0.98\columnwidth]{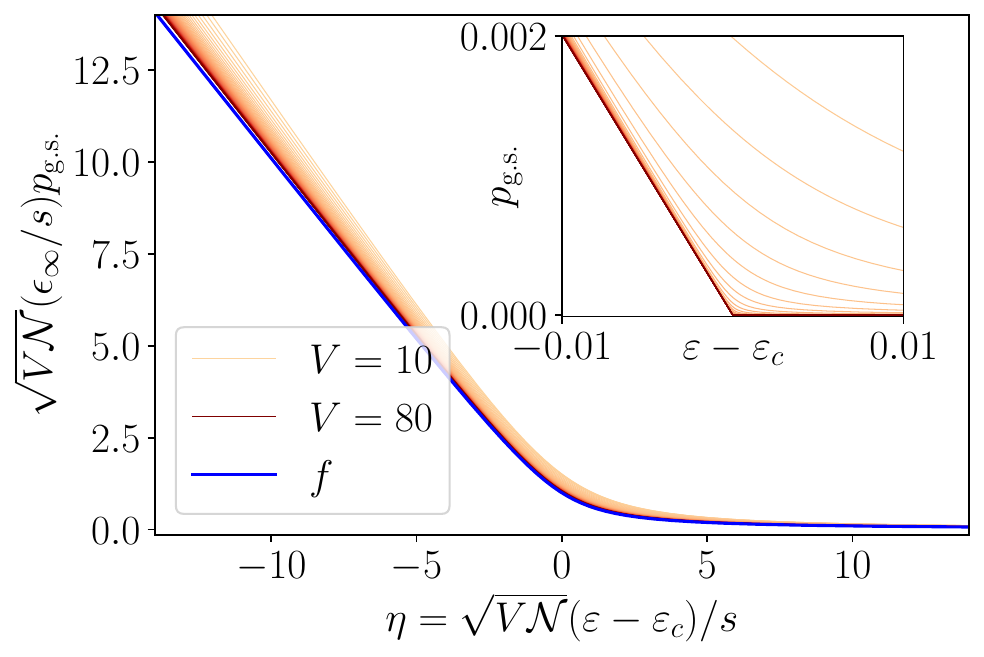}
    \caption{
    \textbf{Scaling collapse of $p_\gs$} for the paramagnetic 1D TFIM with system sizes $V = 10, 11, \dots, 79, 80$.
    $p_\gs$ and $E$, and $E_c$ are computed semianalytically via \eqref{eq:pk-Ek} and \eqref{eq:Ec-sum}.
    The scaling collapse agrees well with the moment-expansion prediction \eqref{eq:scaling}.
    \textbf{Inset:} unrescaled data.
    }
    \label{fig:ising_1d_scaling-collapse}
\end{figure}

\section{Exponential finite-size scaling of transition}
For Gaussian random matrices, the high-temperature phase unambiguously exists in the large-system limit.
In that limit the critical energy densities approach $\varepsilon_{c-} = 1/2$, $\varepsilon_{c+} = 3/2$,
and the phase transitions sharpen as the system grows.

In spin systems the nature of the high-temperature phase is more subtle because the critical regions approach $\varepsilon_\infty$, and each other, like $1/V$.
If the system had a finite size scaling controlled by $V^{-\nu}$ with exponent $\nu < 1$,
the high-temperature phase would not meaningfully exist:
the critical regions $|\varepsilon - \varepsilon_{c\pm}| \sim V^{-\nu}$
would include the whole putative high-temperature phase.

But in fact the condensation transition has exponential finite-size scaling.
A moment expansion (App.~\ref{app:scaling-collapse}), neglecting third and higher moments, gives
\begin{align}\label{eq:scaling}
\begin{split}
    p_{\gs} &= \sqrt{V2^V} \frac s {\varepsilon_\infty}f\Big(\sqrt{V2^V}(\varepsilon - \varepsilon_{c})/s\Big)\;, \\
    f(\eta) &= \frac 1 2 \Big(-\eta + \sqrt{\eta^2 + 4}\,\Big)\;.
\end{split}
\end{align}
As a result of the exponential scaling,
there is an exponential separation of energy scales:
the critical regime $|\varepsilon - \varepsilon_c| \sim 1/ \sqrt{V2^V}$
is much smaller than the high-temperature regime $|\varepsilon - \varepsilon_c| \le O(1/V)$.
This is already visible in Fig.~\ref{fig:fig1}, which shows a region of small $p_{\gs}$ for each Hamiltonian.

Fig.~\ref{fig:ising_1d_scaling-collapse} shows a scaling collapse for average $p_\gs$ computed for the paramagnetic 1D TFIM via \eqref{eq:pk-Ek} (see App.~\ref{app:pgs-Eav-tfim} for details of that calculation and App.~\ref{app:scaling-collapse} for comparison to sampling data).
The scaling collapse agrees with the scaling form \eqref{eq:scaling}, 
up to finite size corrections which we attribute to the higher moments neglected in the derivation of \eqref{eq:scaling}.

\begin{figure}[t!]
    \centering
    \includegraphics[width=0.95\columnwidth]{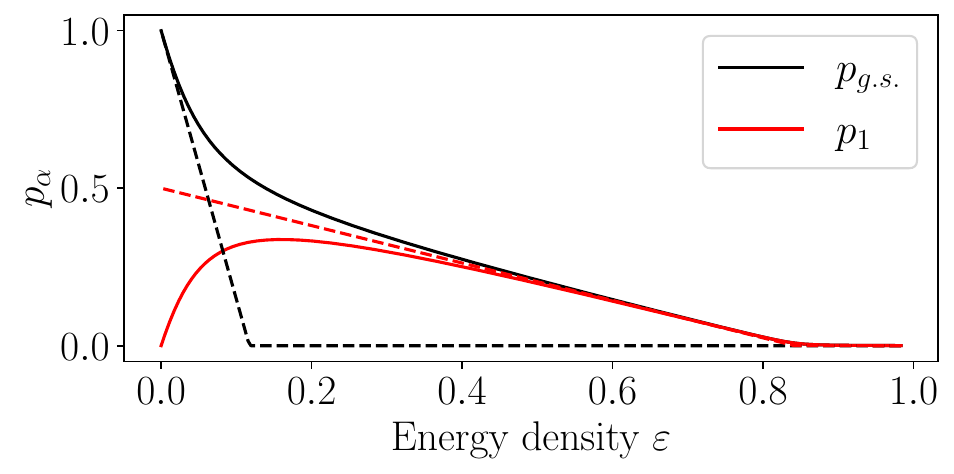}
    \caption{
    \textbf{Ground and first excited state probabilities} for the ferromagnetic 1D TFIM $J = 1$, $h_x = {0.5}$ at $V = 11$.
    At $\varepsilon_{c1} \approx 0.84$ the system undergoes a transition into a regime with $p_\gs = p_1 = 0.5(1 - \varepsilon / \varepsilon_{c1})$, i.e. $O(1)$ weight on the nearly-degenerate ground space.
    At $\varepsilon_{c0} \approx 0.12$ it undergoes a crossover to a regime with $p_\gs \approx 1 - \varepsilon/\varepsilon_{c0}$ and $p_1 \approx \varepsilon / \varepsilon_{c0}$.
    $p_\gs$ and $E$ are calculated semianalytically via Eq.~\eqref{eq:pk-Ek}.
    }
    \label{fig:near-degeneracy}
\end{figure}

\section{Degeneracies and near-degeneracies}
So far we have implicitly assumed that the ground state is non-degenerate.
When the ground state is nearly degenerate, however, the picture is more complicated.
Consider the ferromagnetic phase of the 1D TFIM, which has two $\mathbb Z_2$-symmetric states separated by a gap $\Delta \sim h_x^V$.
If $\Delta \ll E_\infty / \mathcal N $ 
the system undergoes a transition and a crossover:
first, at $E_{c1}^{-1} = \frac 1 {\mathcal N - 2} \sum_{\alpha > 2} E_\alpha^{-1}$,
it develops $O(1)$ weight on the nearly-degenerate ground space, distributed evenly between the two states;
then, at $E_{c0}^{-1} = \frac 1 {\mathcal N - 1} \sum_{\alpha > 1} E_\alpha^{-1} \approx (\mathcal N \Delta)^{-1}$
the weight shifts to the ground state.
Fig.~\ref{fig:near-degeneracy} shows this scenario; there $p_\gs$ and $E$ are again calculated semianalytically as a function of $\beta$ via Eq.~\eqref{eq:pk-Ek}.
If $\Delta \gtrsim E_\infty / \mathcal N$ the intermediate regime, with even weight on the two states, is small.

\section{Conclusion}
A wide variety of spin models display an \textit{eigenstate condensation transition}, in which states drawn from the Haar ensemble conditioned on energy expectation value develop macroscopic weight on their ground states  or anti-ground-states.
We have used numerical methods from the Bayesian statistics community---nested sampling and Galilean Monte Carlo---to confirm the existence of three phases: a ground-state phase, a high-temperature phase, and an anti-ground-state phase.
We have given a compact expression for the critical energies
and an analytical large-system scaling form for the ground state weight in local spin systems;
this scaling form predicts exponential finite size scaling. 
Consequently the critical region shrinks exponentially as the system size grows,
and the high-temperature phase should be understood as an extended phase even in the large-system limit.
Our results depend only on the gross properties of the density of states:
its moments and its behavior near the ground state.
Any system whose density of states goes to zero as $E \to E_\gs, E_{\antigs}$ will display an eigenstate condensation transition from a high-temperature phase to a ground state or anti-ground-state phase;
any system with an extensive free energy %
will display the finite-size scaling we discuss.

The ensemble of energy-constrained random states is  natural, surprising, and potentially useful.
It is natural in that it mirrors the microcanonical ensemble:
it consists of general pure states, rather than energy eigenstates, at an energy expectation value.
But it is surprising in that its properties are markedly different from the microcanonical or canonical ensemble:
for example, it displays an eigenstate condensation transition.
And it is potentially useful in that it may offer a robust mechanism for generating approximate ground states,
which are necessary for quantum phase estimation \cite{kitaevQuantumMeasurementsAbelian1995b,abramsQuantumAlgorithmProviding1999,aspuru-guzikSimulatedQuantumComputation2005a}---a leading candidate for ground state estimation on error-corrected quantum computers---and perhaps suitable input to magic-state or entanglement distillation protocols \cite{bravyiUniversalQuantumComputation2005,bennettPurificationNoisyEntanglement1996}.

Energy-constrained random states are only useful if there is a protocol that generates them.
Such a protocol might come from mixed quantum-classical dynamics.
When a quantum system thermalizes with a classical system,
the result is the canonical analog of the energy-constrained ensemble \cite{parandekarDetailedBalanceEhrenfest2006a};
we expect that the mixed quantum-classical dynamics can arise as a mean-field, semiclassical limit of a purely quantum system.
Such a system would throw into sharp relief other questions about the eigenstate condensation transition.
Can it be understood in terms of symmetry breaking and the renormalization group?
Ground state search is NP-hard, indeed QMA-hard
\cite{kempeComplexityLocalHamiltonian2006},
so it is unlikely that a simple protocol can generate energy-constrained states for all Hamiltonians.
What kinetic obstructions might block relaxation to the energy-constrained ensemble?

\begin{acknowledgements}
We are grateful for helpful conversations with Yuxin Wang, Chris Jarzynski, Daniel Mark, Daniel Ish, and Nicole Yunger Halpern. 
M. W. acknowledges the Institute for Advanced Study and the Goldberger family for supporting his studies. 
N.B. was funded by the United States Naval Research Laboratory basic research base program.
This work was performed while C.D.W. held an NRC Research Associateship award at the United States Naval Research Laboratory;
it was supported by the DOD HPCMPO at the ERDC DSRC.
Our implementation of nested sampling and Galilean Monte Carlo for quantum states is available at
\cite{christopherdavidwhite2Christopherdavidwhite2QuantumNestedSamplingjl2025};
scripts and data used to generate the plots in this paper are available at \cite{white_2026_18509083}.
While revising the manuscript to respond to referee comments
we used Claude (Opus 4.6/4.7 with Claude Code) to optimize code and search literature.
\end{acknowledgements}

\bibliography{references}

\clearpage

\appendix
\appendixsubsectionformat

\section{Nested sampling and Galilean Monte Carlo}

Nested sampling
\cite{skillingNestedSampling2004,%
skillingNestedSamplingGeneral2006}
is a method for evaluating integrals of the form
\begin{align}\label{eq:ns-goal}
    f_\beta = \int d\mu(\psi)\; L(\psi) f(\psi),
\end{align}
where $\mu(\psi)$ is a base measure on some manifold (e.g. the Haar measure on $\mathds {CP}^{\mathcal N-1}$),
$L(\psi)$ is a likelihood, and
$f(\psi)$ some function;
it was designed to be efficient even when $L(\psi)$ is sharply peaked and
the integral is dominated by a small volume.
Nested sampling originated in the Bayesian statistics community
as a method for evaluating evidence integrals;
there $L$ is the Bayesian likelihood, $f(\psi) = 1$.
It has since been applied to calculate thermodynamic integrals \cite{ashtonNestedSamplingPhysical2022},
where $L(\psi) = e^{-\beta E(\psi)}$, $E$ is an energy, and $\beta$ is the inverse temperature.
The central feature of nested sampling is that it creates a series of
samples from a uniform distribution over $\mu$ that become progressively concentrated
in regions of increasingly large $L$, i.e.\ low $E$.
In this work we take advantage of nested sampling's production of uniform samples,
but we do not use these samples to evaluate integrals.

\begin{figure}[t]
    \centering
    \includegraphics[width=0.97\columnwidth]{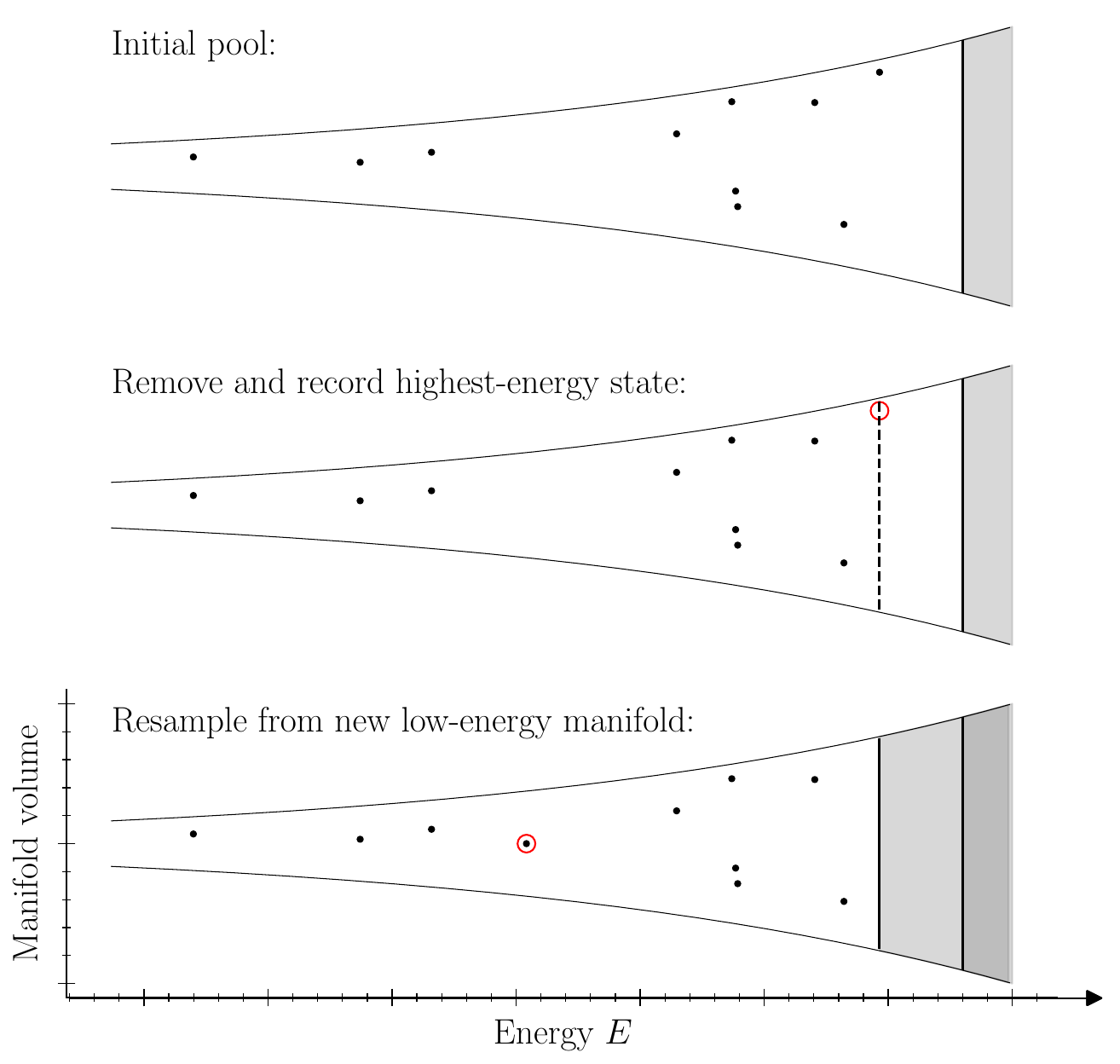}
    \caption{\textbf{Nested sampling} iteratively removes and records the highest-energy state, then resamples from the submanifold with energy below that previous highest energy; the measure of the constrained manifold shrinks exponentially.
    Nested sampling requires an efficient resampling procedure.
    Frequently the resampling procedure duplicates an existing point and moves it at random through the low-energy submanifold;
    we do so using \textit{\gmc}.
    }
    \label{fig:ns}
\end{figure}

Nested sampling works iteratively on a set of $\nlive$ states, called \textit{live points}.
At each iteration one removes and records the highest-energy live point;
call the live point removed and recorded at iteration $i$ $\ket{\psi^*_i}$ and its energy $E^*_i = \braket{\psi^*_i|H|\psi^*_i}$.
One then draws a replacement uniformly from the submanifold%
\begin{align}
  \{\ket{\psi} : \braket{\psi| H| \psi} < E_i\}\;.
\end{align}
(We use \gmc, described below, to draw the replacement.)
At each step, then, the live points are uniformly distributed subject to the constraint $\braket{\psi | H | \psi } < E^*_i$
without any further equilibration or sampling.
The live points are therefore always distributed according to the correct (uniform) distribution.

The output of nested sampling is the ordered list of states $\ket {\psi_i^*}$ removed and recorded and their energies $E_i^*$.
The energies have convenient measure-related properties,
but for our purposes the state properties are more important.
Because the pool states after step $i$ are uniformly distributed in $\braket{\psi | H| \psi } < E^*_i$,
each recorded state $\ket{\psi_i^*}$ is sampled uniformly $\braket{\psi | H| \psi } = E^*_i$:
precisely the manifold of constrained-energy states we wish to sample.

\begin{figure}
    \centering
    \includegraphics[width=0.90\columnwidth]{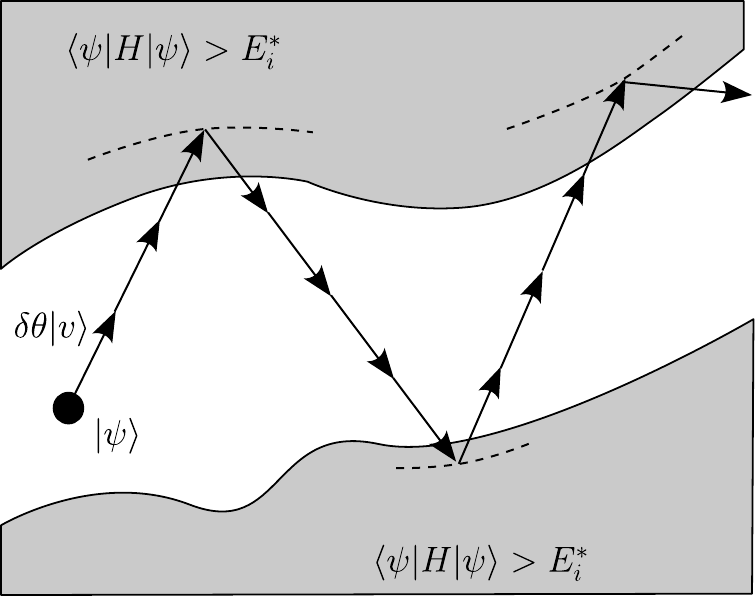}
    \caption{\textbf{\gmc}: the state $\psi$ makes small (size-$\theta$) steps along a randomly chosen vector $\ket v$ until it leaves the high-energy subspace; the vector is then specularly reflected through the energy isosurface. Because the vector is specularly reflected, the component parallel to the energy isosurfaces is preserved and the state ``advects'' through the low-energy submanifold, rather than ``diffusing'' as it would in slice sampling.  }
    \label{fig:gmc}
\end{figure}

The efficiency and accuracy of nested sampling depend on the efficiency and accuracy of the resampling procedure.
We resample by \textit{Galilean Monte Carlo},
which duplicates a live point and moves it through the constrained low-energy manifold,
specularly reflecting from the constraint boundary.
To understand Galilean Monte Carlo, start with  a conventional Monte Carlo step.
Heuristically, one duplicates one of the live points---call it $\ket \psi$---and repeatedly moves it a small amount along a vector $\ket{v}$.
If the result of the move is in the constrained manifold one accepts it;
if not, one rejects it and tries again.%
Typically many such moves are required to make the correlations of the new point sufficiently small.
These Monte Carlo steps, as applied to quantum states, can be understood in terms of unitaries.
The Monte Carlo step consists of
\begin{align}\label{eq:exp-map}
    \ket{\psi} \mapsto e^{i\theta \ket{v}\bra{\psi}} \ket \psi
\end{align}
for some small $\theta$, with $\ket v$ chosen randomly from the Haar ensemble.

For each step to have a significant probability of acceptance $\theta$ must be small.
This is because the measure of the constrained-energy manifold is exponentially small in the number of nested sampling steps
and the two-dimensional subspace $\mathcal H_{\psi,v}$ is (in part) a random subspace,
so the set of $\theta$ for which the energy constraint is satisfied cannot be large.
Since one desires the final state to be uncorrelated with the initial state,
one must repeat many times with many random vectors $\ket{v}$.
The state $\ket{\psi}$ follows a random walk through the low-energy submanifold,
so it explores that manifold diffusively.

Galilean Monte Carlo \cite{betancourtNestedSamplingConstrained2011a,skillingBayesianComputationBig2012} (Fig.~\ref{fig:gmc}, Alg.~\ref{alg:gmc}) replaces diffusion with advection by specularly reflecting from the constraint.
In Galilean Monte Carlo, as applied to quantum states, one first takes many steps in a single direction with a small stepsize $\theta$;
after the first step that takes one outside the low-energy manifold,
one specularly reflects and continues with the new (reflected) vector.
(Note that the tangent vector along which the point is moved must itself undergo parallel transport $\ket v \leftarrow e^{-iF\theta} \ket v$, where
$F = \ketbra{\psi}{v} + \ketbra{v}{\psi}$.)
The specular reflection can be computed by a Householder reflection about $\partial_\psi \braket{\psi | H \psi}$; some care is required, since $\ket \psi$ and $\ket v$ are both complex.
One continues in this way until the state has traveled some total distance $L$.
The state returned by Algorithm 1 may be outside the low-energy manifold.
In that case---as in conventional Monte Carlo---one rejects the state and tries again, choosing a new random direction $\ket v$.
The rejection rate can be made small by making the step size $\theta$ small.
We make multiple \gmc{} moves, with new random directions for each move, and check for convergence in the number of such moves.
The component of the initial $\ket v$ that is parallel to the boundary of the
allowed region is preserved by the reflection, and as a result the position changes
in a correlated manner (i.e., advects) through all steps in the trajectory.
The process is still diffusive over times longer than a single
\gmc{} trajectory, but with a larger effective step length
set by the advection distance.

\begin{figure}[t!]
\begin{algorithm}[H]
\caption{\gmc}
\label{alg:gmc}
\begin{algorithmic}[1]
    \Procedure {GalileanMonteCarlo}{$H$, $\ket \psi$, $ \theta$, $L$}
    \State Choose $\ket v$ at random such that $\braket{\psi | v} = 0$
    \State $\ell = 0$
    \While{$\ell \le L$}
    \While{$\ell \le L$ and $\braket{\psi | H | \psi} < E_*$ }
    \State $F \gets \ketbra{\psi}{v} + \ketbra{v}{\psi}$
    \State $\ket \psi \gets e^{-iF\theta}\ket \psi$
    \State $\ket v \gets e^{-iF\theta}\ket v$
    \State $\ell \gets \ell +  \theta$
    \EndWhile
    \State $v \gets$ reflection of $v$ about energy isosurface at $\ket \psi$
    \EndWhile
    \State \Return $\ket \psi$
    \EndProcedure
\end{algorithmic}
\end{algorithm}
\end{figure}

Nested sampling with \gmc{} has a number of limitations. 
The number of nested sampling steps required to reach any fixed energy is exponential in system size (see App.~\ref{app:ns-difficulty}).
Each \gmc{} step requires many projections onto 2D subspaces, one for each specular reflection; these are again exponentially difficult in system size.
(Moreover we expect the number of specular reflections required to reach a fixed path length to grow as the energy decreases.)
And the whole process is most easily implemented in the energy eigenbasis.
That is because the process requires many matrix-vector multiplications of the form $H \ket v$;
such multiplications must be at least a factor of $V$ slower in the computational basis than in the diagonal basis,
since $H$ has $O(V2^V)$ elements in the computational basis compared to $O(2^V)$ in the diagonal basis.
We find that in practice Julia's sparse matrix vector multiplication is some constant factor slower still.
Consequently, nested sampling with \gmc{} is (for the present) limited to exact-diagonalization-scale systems $V \lesssim 14$.

\section{Energy distribution}\label{app:ed}
To derive an analytical estimate of the energy distribution of the energy-constrained random state, we follow a standard derivation of the Boltzmann distribution \cite{hillIntroductionStatisticalThermodynamics1986}.
Consider many sample sets of $\mathcal M$ random states $\ket{\psi^{(k)}}$, $k = 1, \dots, \mathcal M$.
We seek the most probable sample average weight on energy eigenvector $\ket{E_\alpha}$.
For each sample set that weight is
\begin{align}
    p_\alpha = \sum_{k = 1}^{\mathcal M} |\braket{E_\alpha|\psi^{(k)}}|^2\;.
\end{align}
Ignore the normalization and energy constraints---we will re-introduce these via Lagrange multipliers.
Each sample set is a vector with $2\mathcal M\mathcal N$ real components $\mathrm{Re},\,\mathrm{Im} \braket{E_\alpha|\psi}$.
The projection of that vector onto the $2\mathcal M$-dimensional subspace of components corresponding to a particular $\alpha$ lies in a spherical shell of radius $\sqrt{p_\alpha}$.
Because we have dropped the constraints, the components corresponding to different energy eigenstates $\ket{E_\alpha}$ are independent,
and the volume of a collection of sample sets with a particular $p_\alpha$ is the volume of that radius-$\sqrt{p_\alpha}$ spherical shell in that $2\mathcal M$-dimensional (real) space;
this volume is $\Omega \propto p_\alpha^{\mathcal M }$.
Maximizing $\log \Omega$ with Lagrange multipliers for normalization and energy for a typical sample-average weight on eigenstate $\ket \alpha$ gives the form \eqref{eq:pk-Ek}.
We give some details of the maximization, as well as the many non-rigorous steps in our argument, in App.~\ref{app:weight-deriv}.

\section{Additional details of numerical methods}\label{app:numerical-methods}

In this section we give more detail on our numerical methods. 
In App.~\ref{ss:ns} we give some measure-related and complexity-theoretic properties of nested sampling,
and in App.~\ref{ss:convergence} we describe convergence testing in the Galilean Monte Carlo parameters.
In App.~\ref{app:Mneg1} we describe how we compute the first negative moments of the densities of states, hence the critical energies.

\subsection{Nested sampling with \gmc}\label{ss:ns}
\subsubsection{Measure properties of nested sampling}

Although we do not use the measure-related properties of nested sampling,
it is useful to review them.
(See \cite{ashtonNestedSamplingPhysical2022} for a more detailed review.)
The central result is that each sampling step reduces the measure of the constrained manifold $E < E_*$ by a factor \(\approx 1 - 1/\nlive\) at each step.
{ (In our case, the measure in question is the Haar measure.)
Consequently one can associate a measure estimate with each sample, and estimate integrals with respect to that measure.}

To see the measure, write
\begin{align}
  X_i = \haarint{\psi} [\braket{\psi|H|\psi} < E^*_i]
\end{align}
for the volume of the submanifold with energy $< E^*_i$. (Here $[\cdot]$ is the Iverson bracket: 1 if the argument is true; 0 if false.)
Then because the live points are uniformly distributed in the constrained manifold, the next volume $X_{i+1}$ is
\begin{align}
  X_{i+1} = t_i X_i
\end{align}
where the $t_i$ are beta distributed
\begin{align}
  P(t_i) = \nlive t_i^{\nlive - 1}
\end{align}
Consequently
\begin{align}
  X_{i} = \bar X_i[1 + O(\nlive^{-1}\sqrt{i})]\;
\end{align}
where the $\bar X_i$ are estimators
\begin{align}
  \bar X_i = t^i\;,
\end{align}
writing
\begin{align}
    t = (1 - 1/\nlive)^i\;.
\end{align}

The measure property allows one to estimate Haar integrals:
\begin{align}
  \label{eq:NS-Haar}
  \begin{split}
  g_\beta &= \haarint{\psi} e^{-\beta E(\psi)} g(\psi)\\
  &\approx \sum e^{-\beta E^*_i} g(\psi_i^*) \cdot \frac 1 2 (t^{i-1} - t^{i+1})\;.
  \end{split}
\end{align}
This can be quickly re-evaluated for any $\beta$; indeed, derivatives can be taken semi-analytically.
(This quadrature gives better precision than na\"ive Monte Carlo in the dominant low-energy region.)

Because the unknown true \(X_i\) vary from the estimator \(\bar X_i\), this nested sampling estimate incurs an additional error.
But that additional error cost buys two advantages: improved sampling of the low-energy subspace,
and---when \(f_\beta\) depends solely on the energy---%
the ability to cheaply compute \(f_\beta\) and its derivatives for many \(\beta\).

\subsubsection{Exponential complexity of nested sampling for energy-constrained random states}\label{app:ns-difficulty}

How many resampling steps does it take to reach some energy $E$?
This depends on the probability distribution of $E$ for Haar states:
if the energy CDF is
\begin{align}
    F(E) = \haarint{\psi}\Big[\braket{\psi |H|\psi} < E\Big]
\end{align}
then the measure properties give that the the number of steps $n$ is
\begin{align}
    t^n \sim F(E) \;.
\end{align}
where once again $t = (1 - 1/\nlive)$.
To estimate the CDF,
use that the energy of a random state $\ket \psi$ on a Hilbert space of dimension $\mathcal N$ is very nearly the average of $\mathcal N$ random variables
\cite{auerbachInteractingElectronsQuantum1998,camposvenutiProbabilityDensityQuantum2013,reimannFullExpectationvalueStatistics2019}.%
\footnote{
\cite{reimannFullExpectationvalueStatistics2019} points out that the energy distribution has strongly non-Gaussian tails.
We expect those tails to modify the estimate \eqref{eq:ns-steps}, but not the conclusion that any finite $E$ requires an exponential number of steps.
}
The energy is
\begin{align}
 \braket{\psi | H | \psi} = \frac 1 {\mathcal N} \sum_\alpha \left(\mathcal N |\braket{\psi|\alpha}|^2\right) E_\alpha\;,
\end{align}
where $\ket \alpha$ is the eigenstate of $H$ with energy $E_\alpha$.
Since $\ket{\psi}$ is random, we have $\mathcal N |\braket{\psi|\alpha}|^2 \sim 1$.
The variance of $\left(\mathcal N |\braket{\psi|\alpha}|^2\right) E_\alpha$ is therefore $\sim s^2 V$,
since $s^2V$ is the variance of the density of states.
Consequently the energy is approximately Gaussian with variance $\sim s^2 V 2^{-V}$
and
\begin{align}
    t^n \approx \frac 1 2 \left[1 + \erf\left((E - E_\infty) / 2\sqrt{s^2 V 2^{-V}}\right)\right]\;.
\end{align}
Apply the asymptotic expansion of erfc, take logs, and discard subleading terms to get
\begin{align}\label{eq:ns-steps}
    n \sim \frac 1 {\ln t} \frac {(E-E_\infty)^2}{s^2} V^{-1} 2^V\;.
\end{align}
(Recall that $t$ depends only on the number of live points;
we can take $\ln t$ to be $O(1)$.)
That is, it takes an exponential-in-system-size number of steps to achieve any finite distance $E - E_\infty$ from the infinite-temperature average energy $E_\infty = 2^{-V} \tr H$.

\begin{figure*}[t]
    \centering
    \includegraphics[width=0.45\textwidth]{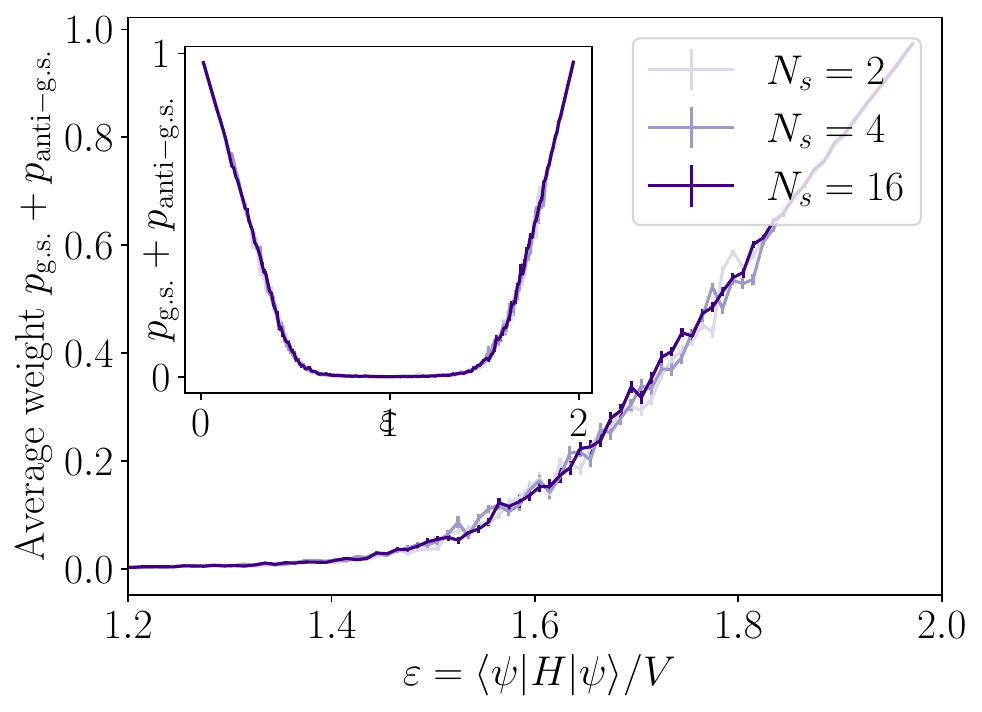}
    \includegraphics[width=0.45\textwidth]{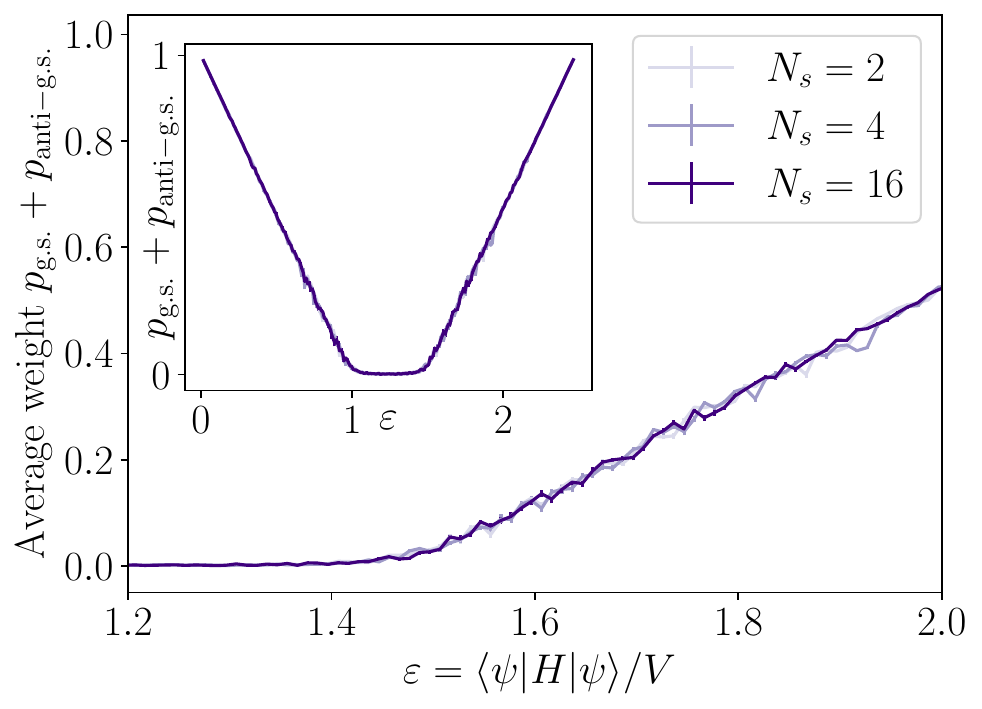}
    \caption{\textbf{Convergence in number of \gmc{} steps per nested sampling step} 
    for $V = 10$ spins in a GOE Hamiltonian (\textbf{left}) and the 1D critical TFIM (\textbf{right}).
    Error bars show standard error of the bin average.
    }
    \label{fig:gmc-convergence}
\end{figure*}

\subsection{Nested sampling with \gmc: simulation parameters and convergence} \label{ss:convergence}

We use nested sampling with $n_\live = {2}$ live points, a \gmc{} step pathlength $L = {32}$, $\delta \theta = {2^{-10}}$, and $N_s = {1}$ to ${16}$ \gmc{} steps per nested sampling step.
We use Haar-random states for the initial pool of live points.
We apply the procedure to $H$ to sample states with $E \lesssim \mathcal N^{-1} \tr H$, and to $-H$ to sample states with $E \gtrsim \mathcal N^{-1} \tr H$.
The result is a list of paired energy densities and states.
We compute $p_\gs + p_\antigs$ for each state;
we average $p_\gs + p_\antigs$ over states in energy density bins of bin width $\delta = {0.01}$ (chosen to balance noise against energy resolution).

If the total distance covered by the \gmc{} steps in each nested sampling step is too small,
then successive samples will be correlated.
This could have a number of consequences.
Most seriously, our estimates may be biased, because we do not adequately sample the space.
Additionally, samples in each bin would be correlated, so  the bin-to-bin variation would be larger 
and our errorbars---which show the standard error of $p_\gs + p_\antigs$---would underestimate the error.

To check that we use enough \gmc{} steps we sample at $N_s = 2, 4, 16$ 
(see Fig.~\ref{fig:gmc-convergence} for a GOE Hamiltonian and the 1D critical TFIM).
All three broadly agree, indicating that even at $N_s = 2$ we are adequately sampling the space and our estimates are not biased.
For $N_s = 2$ and to a lesser extent $N_s = 4$ the bin-to-bin variation is much larger than the errorbars, indicating that successive states are correlated;
for $N_s = 16$ the errorbars are of the same order as the bin-to-bin variation,
indicating that sample-to-sample correlations are negligible.
(In fact the errorbars are still smaller than the bin-to-bin variation by perhaps a factor of 1.5 or 2, estimated by eye, indicating that some correlations remain:
that is, that there is a ``correlation length'' $\sim 2-4$ states.
We believe this to be an acceptable level of convergence, especially given the broad agreement of the three  $N_s$.)

Our convergence requirements are much less stringent than the convergence requirements for estimating integrals.
In estimating an integral, any residual correlations will systematically bias the measure---whereas in our case they merely make our estimates more noisy.
Moreover we use an exceptionally small number of live points $n_\live = 2$;
in estimating an integral, this would almost certainly result in unacceptable statistical noise in the measure estimate.

\begin{figure*}[t]
    \centering
    \includegraphics[width=0.32\textwidth]{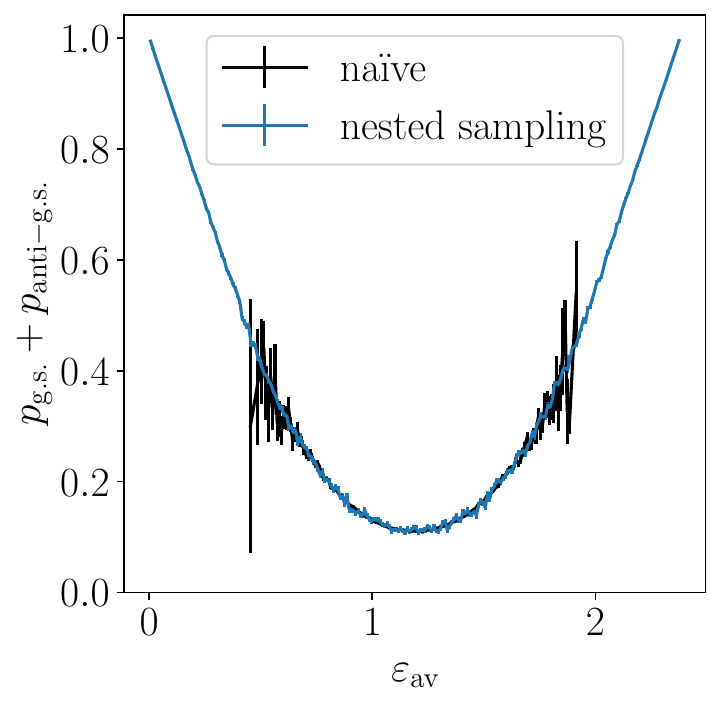}
    \includegraphics[width=0.32\textwidth]{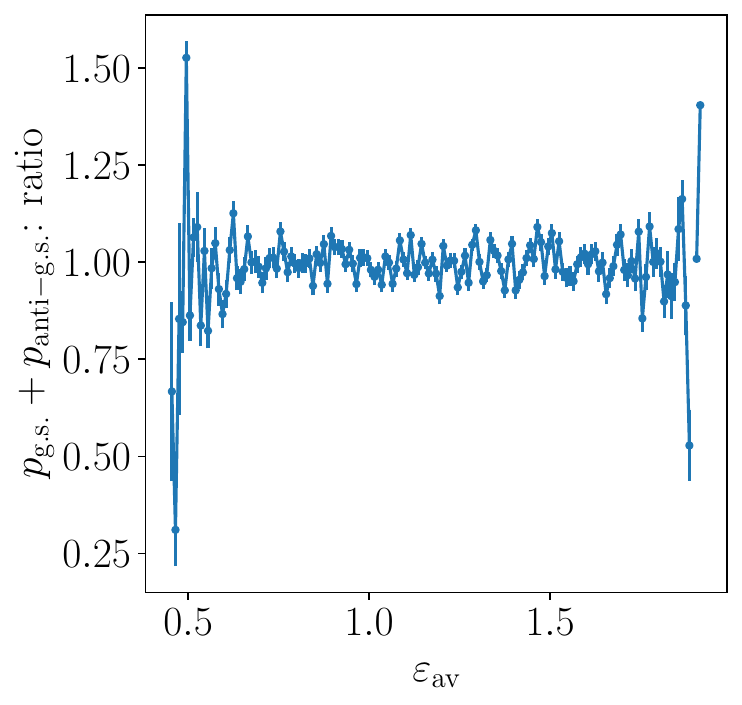}
    \includegraphics[width=0.32\textwidth]{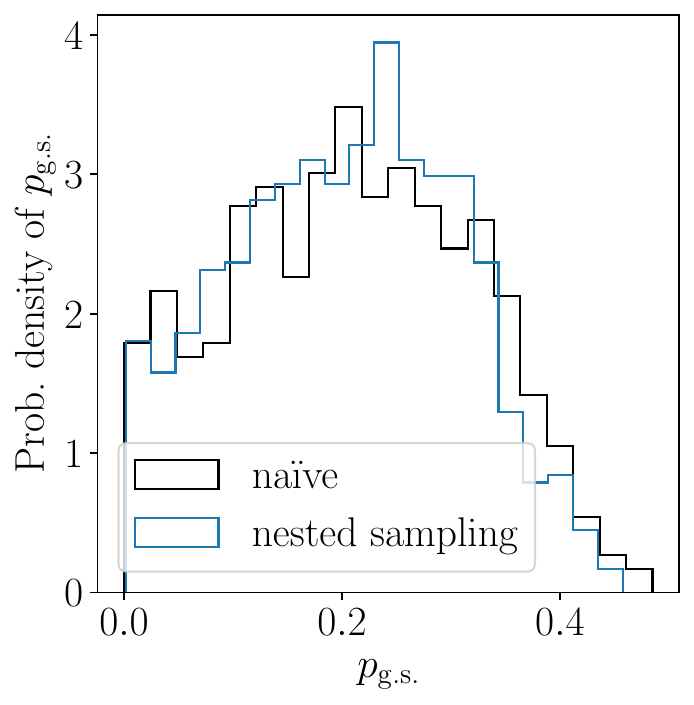}
    \caption{
        \textbf{Comparison of nested sampling with na\"ive sampling from the Haar ensemble} for the $L_x = 4$ critical TFIM 
        For nested sampling we use $n_\live = 2048$ live points, $N_s = 16$, and other parameters as described in the beginning of the section.
        For the na\"ive sampling we take $10^7$ states i.i.d. from the Haar distribution.
        \textbf{Left:} As in main text Fig.~1 and elsewhere, $p_\gs + p_\antigs$ averaged over energy density bins of width $\delta = 0.01$.
        The two agree near the middle of the spectrum;
        as the energy density increases or decreases, the na\"ive Haar sampling becomes unreliable
        because there are too few states in any bin.
        \textbf{Center:} ratio $[\text{na\"ive}] / [\text{nested}]$ of bin-averaged $p_\gs + p_\antigs$ with bin width $\delta = 0.01$.
        Error bars are the combined standard errors of the two.
        Again the two agree within statistical error, which grows away from mid-spectrum.
        \textbf{Right:} histogram of $p_\gs$ for a bin of width $\delta = 0.01$ around $\varepsilon_0 = 0.75$,
        i.e. $0.745 < \varepsilon_\av \le 0.755$.
        Na\"ive sampling nested sampling agree by eye.
    }
    \label{fig:ns-naive}
\end{figure*}

To further validate nested sampling in our parameter regime,
we consider small systems.
First we consider a large $\nlive$ on a very small system $V = 4$, and compare to na\"ive sampling of i.i.d. Haar states,\footnote{
    By ``na\"ive sampling'' we mean constructing normalized Haar-random states
    \begin{align}
    \begin{split}
    \psi_\alpha &\sim \mathcal N(0,1;\mathds C) \\
    \psi_\alpha &\leftarrow \psi_\alpha / \|\psi\|
    \end{split}
    \end{align}
    and binning the results by energy.
}
to confirm that nested sampling gives the correct $p_\gs$ in the large-$\nlive$, small-$V$ limit.
Fig.~\ref{fig:ns-naive} compares na\"ive sampling of $10^7$ states to nested sampling with $\nlive = 2048$;
the two agree within statistical error.

That agreement is only good in the limit of small energy windows $\delta$.
For larger energy windows, the binning procedure interacts with nested sampling to produce a small bias $\propto \delta$.
Nested sampling produces samples with a density that grows as one approaches the edge of the spectrum,
so the bin-average of those samples will be weighted toward the outer edge of the bin.
The Haar distribution, by contrast is much denser near mid-spectrum,
so the bin-average of Haar samples will be weighted toward the inner edge of the bin.
The resulting discrepancy is $\lesssim \delta \frac{\partial }{\partial \varepsilon_\av} p_{\gs, \antigs}\lesssim \delta$,
since for our models $\frac{\partial }{\partial \varepsilon_\av}p_{\gs, \antigs} \lesssim 1$.
For larger $\delta$ the discrepancy is visible in histograms like (the analog of) Fig.~\ref{fig:ns-naive} right,
but for our $\delta = 0.01$ the discrepancy is small.
If that were not the case, or if we required high precision, we could correct the bias by performing integrals as in Eq.~\eqref{eq:NS-Haar}%
---as it is, we do not deem that necessary.

\begin{figure*}[t]
    \includegraphics[width=0.28\textwidth]{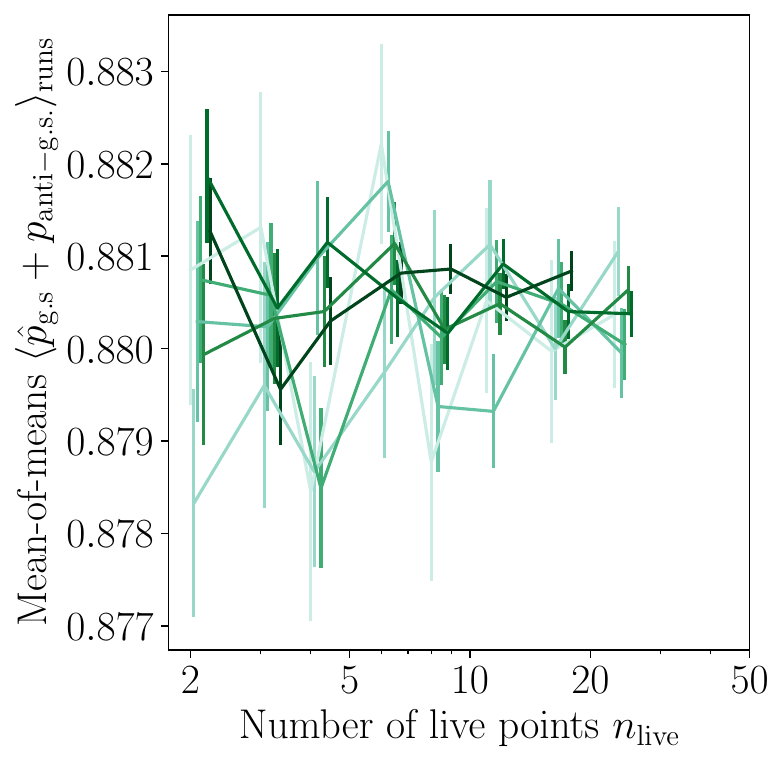}
    \includegraphics[width=0.26\textwidth]{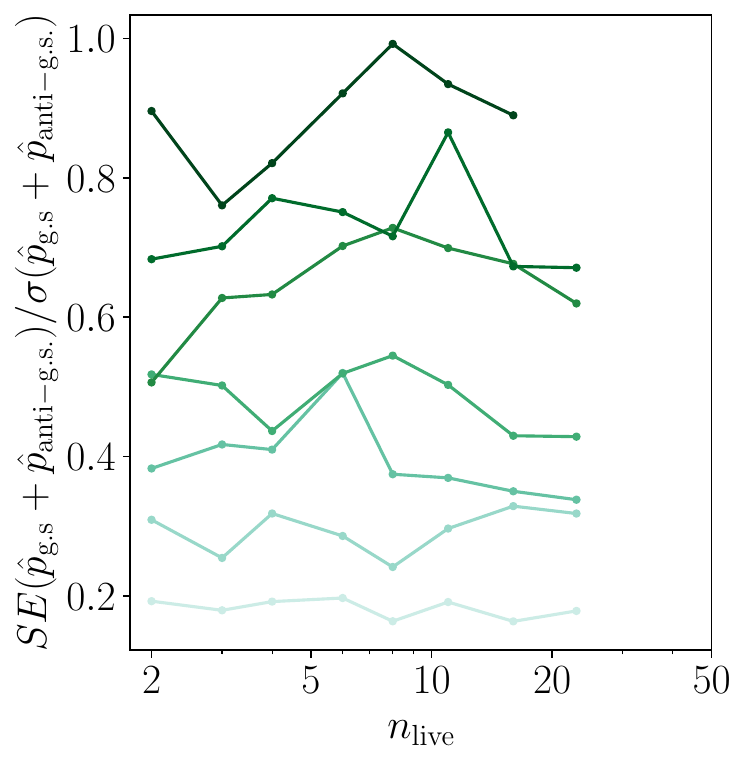}
    \includegraphics[width=0.40\textwidth]{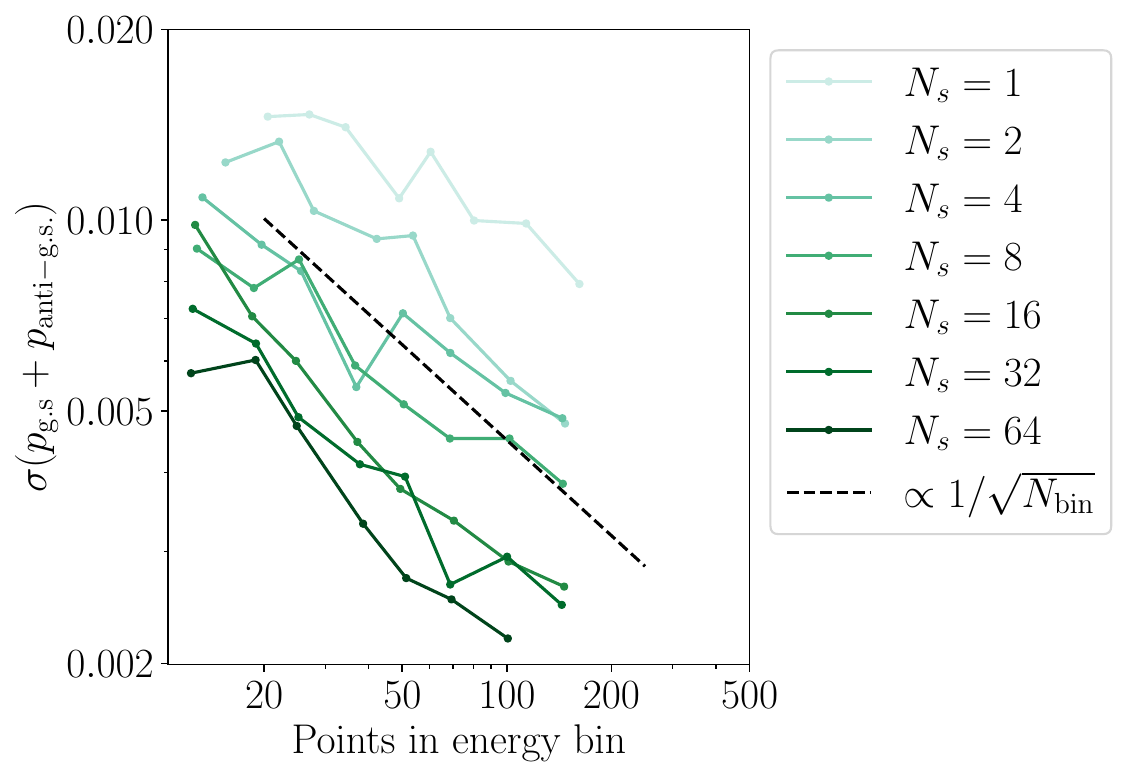}
    \caption{
    \textbf{Nested sampling convergence in $\nlive$} for a $V = 6$ critical Ising Hamiltonian in an energy density window $0.295 < \varepsilon_\av \le 0.305$.
    \textbf{Left:} average and standard error over runs of the per-run estimator $\hat p_\gs + \hat p_\antigs$. The average does not display any evident drift as a function of the number of live points $n_\live$, across more than an order of magnitude (from 2 to 64).
    \textbf{Center:} Per-run standard error, averaged over runs,
    over the standard deviation between runs of the estimator $\hat p_\gs + \hat p_\antigs$.
    For small number of GMC walks $N_s$, the per-run standard error underestimates the uncertainty due to correlations between samples.
    \textbf{Right:} Standard deviation between runs of the estimator, against the number of states in the energy bin.
    This displays the expected central limit theorem scaling.
    }
    \label{fig:nlive}
\end{figure*}

We then consider moderately small systems $V = 6$ purely in nested sampling.
We run nested sampling 100 times at each of a large number of $\nlive$ and \gmc{} steps $N_s$.
For each nested sampling run we compute the per-run ground state probability estimator
\begin{align}
    \hat p_\gs(E;\delta) =
    \frac 1 {n_\bin} \sum_{E - \delta / 2 < E(\psi_\alpha) < E + \delta / 2} | \langle \gs | \psi^{(\alpha)} \rangle|^2
\end{align}
averaged over states in a bin of width $\delta$ around energy $E$,
and the corresponding per-run standard error
\begin{widetext}
\begin{align}
    \mathrm{SE}(\hat p_\gs) = 
    \frac 1 {n_{\bin}-1} \left[ \sum_{E - \delta / 2 < E(\psi_\alpha) < E + \delta / 2}
    \Big( | \langle \gs | \psi^{(\alpha)} \rangle|^2 - \hat p_\gs(E;\delta) \Big)^2 \right]^{1/2}\;.
\end{align}
\end{widetext}
(In each case $n_\bin$ is the number of points in the bin.)
The per-run standard error $\mathrm{SE}(\hat p_\gs)$ is the run-internal estimate of the error that we use for error bars in our plots, e.g. Fig.~\ref{fig:gmc-convergence};
the average over runs of the standard error
\begin{align}
    \expct{ \mathrm{SE}(\hat p_\gs) }_{\text{runs}}
\end{align}
is (broadly speaking) the typical internal estimate of the error.
We also compute the average and standard deviation over runs of the estimators 
\begin{align}
\begin{split}
    &\langle \hat p_{\gs} \rangle_{\text{runs}} \\
    &\sigma(\hat p_{\gs}) = \langle \hat p_\gs^2 \rangle_{\text{runs}} - \expct{\hat p_\gs}_{\text{runs}}^2\;.
\end{split}
\end{align}
The latter is the true (statistical) uncertainty.
We also compute the analogs of these quantities for the anti-ground-state.

We plot these quantities in Fig.~\ref{fig:nlive}. 
The estimator $\hat p_\gs$ displays the expected central limiting behavior,
regardless of system size.
The per-run standard error $\mathrm{SE}(\hat p_\gs)$ underestimates the statistical uncertainty by an $O(1)$ factor for a small number $N_s$ of Galilean Monte Carlo steps $N_s$,
but for $N_s = 64$ the error is small.
This is broadly consistent with Fig.~\ref{fig:gmc-convergence},
though the number of \gmc{} steps required to decorrelate the samples is somewhat larger than one would expect from Fig.~\ref{fig:gmc-convergence}.
This is because Fig.~\ref{fig:nlive} uses a bin near enough to the edge of the spectrum that the error bars would barely be visible in the analogue of Fig.~\ref{fig:gmc-convergence}.
Indeed for higher energies, nearer the middle of the spectrum, plots analogous to \ref{fig:nlive} center show that $\mathrm{SE}(\hat p_\gs)$ converges to the true uncertainty appreciably more quickly with $N_s$.

\subsection{Evaluating the first negative moment} \label{app:Mneg1}

The critical energy is given by the first negative moment of the bulk density of states
\begin{align}\label{app:eq:Mneg1}
    E_c^{-1} = \frac 1 {\mathcal N_\bulk} \sum_{\alpha \ne \gs} \frac 1 {E_\alpha} = \frac 1 {\mathcal N_\bulk} \tr H_\bulk^{-1}
\end{align}
For GOE and GUE systems we use the tridiagonal form (Sec.~\ref{app:GbE}) to compute this exactly for systems up to $V \le 16$.
For nonintegrable and Bethe ansatz integrable spin systems we use stochastic traces, limited to small system sizes $V \lesssim 25$; 
for free fermion integrable Hamiltonians \eqref{app:eq:Mneg1} can be evaluated semianalytically for large systems.

\subsubsection{Non-free-fermion Hamiltonians}\label{app:Ec-stoch}

\textit{Stochastic trace} methods estimate
\begin{align}
   \tr f(A)  
\end{align}
given a multiplication primitive
\begin{align}
    f_A(v) : \ket v \mapsto f(A) \ket v
\end{align}
by applying $f(A)$ to many random $\ket v$.
The simplest stochastic trace method is the Girard-Hutchinson stochastic trace \cite{girardFastMonteCarloCrossvalidation1989,hutchinsonStochasticEstimatorTrace1989},
which averages $\braket{v| f(A)| v}$ over many random $v$.
We use a Girard-Hutchinson variant in which the $\ket v$ have i.i.d. real Gaussian entries in the computational basis.

In our case the multiplication primitive is the matrix left division
\begin{align}
    \ket{v} \mapsto H_\bulk^{-1} \ket{v} = \big[H(1 - \ketbra{\gs}{\gs})\big]^{-1} \ket v\;.
\end{align}
$H$ is sparse, so multiplication by $H_\bulk$
\begin{align}
    H_\bulk \ket v = H(1 - \ketbra{\gs}{\gs}) \ket v
\end{align}
can be efficiently computed.
We use that multiplication in a Krylov conjugate gradient method \verb+cg()+ from Krylov.jl \cite{montoison-orban-2023} to compute $\ket{v} \mapsto H_\bulk^{-1} \ket{v}$.

\begin{figure}[t]
    \centering
    \includegraphics[width=0.98\columnwidth]{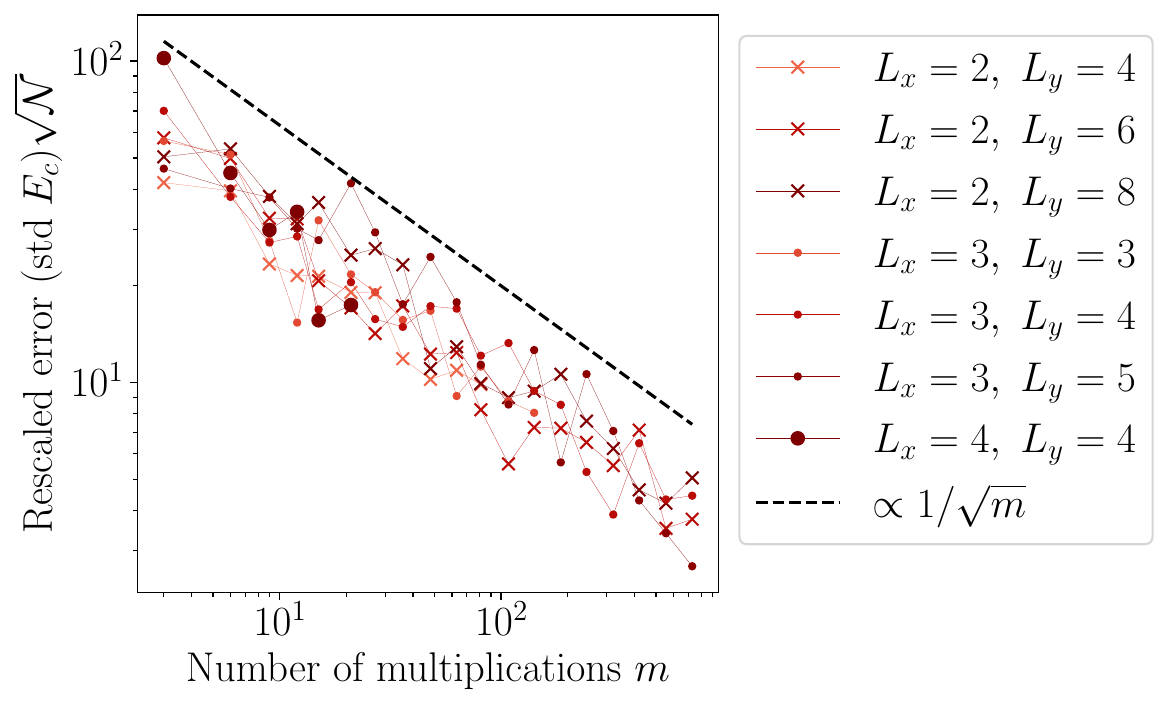}
    \caption{\textbf{Stochastic trace convergence for the 2D TFIM:} standard deviation of 200 estimates of the critical energy, each using $m$ matrix multiplications. }
    \label{fig:Mneg1_ising_convergence}
\end{figure}

Fig.~\ref{fig:Mneg1_ising_convergence} shows the standard deviation in the stochastic trace estimate of $E_c$ as a function of the number of multiplications $m$.
We find that the error is
\begin{align}
    \mathrm{std}\, E_c \lesssim \frac{200}{\sqrt{m\mathcal N}}\;.
\end{align}
We use this to guide our choice of $m$ for the figures in the main text.
In each case we then validate with nonparametric bootstrap;
the resulting errorbars are smaller than the dots.

\subsubsection{Free fermion integrable Hamiltonians}\label{app:Ec-ff}
Free fermion integrable Hamiltonians can be written
\begin{align}\label{eq:ff}
    H = \sum_{k=1}^V \epsilon_k n_k
\end{align}
where the $n_k$ are occupation number operators.
(The 1D TFIM can be put in this form by Jordan-Wigner transformation, on which more below.
The index $k$ is not necessarily a momentum.)
To evaluate
\begin{align}
    E_c^{-1} = \frac 1 {\mathcal N - \mathcal N_\gs} \sum_{\alpha \ne g.s.} \frac 1 {E_\alpha}\;,
\end{align}
where $\mathcal N_\gs$ is the ground state degeneracy,
we use the identity
\begin{align}
\begin{split}
\sum_{\alpha \ne g.s.} \frac 1 {E_\alpha} &= \tr H_\bulk^{-1}\\
&= \int_0^\infty d\lambda\; \tr e^{-\lambda H_\bulk} \\
&= \int_0^\infty d\lambda\; \Big(\tr \left[e^{-\lambda H}\right] - \mathcal N_\gs\Big)
\end{split}
\end{align}
where $H_\bulk$ is the projection of the Hamiltonian onto the subspace perpendicular to the degenerate ground space.
(Recall that we have shifted the Hamiltonian so the ground state energy is $E_\gs = 0$.)

Given the single-fermion energies, the integrand can be computed in time $O(V)$ via
\begin{align}
    \tr e^{-\lambda H} = \prod_k \left(1 + e^{-\lambda \epsilon_k}\right)\;;
\end{align}
we evaluate the integral numerically by Gauss-Kronrod quadrature (QuadGK.jl \cite{JuliaMathQuadGKjl2025}).

\begin{figure*}[t]
    \centering
    \includegraphics[width=0.98\textwidth]{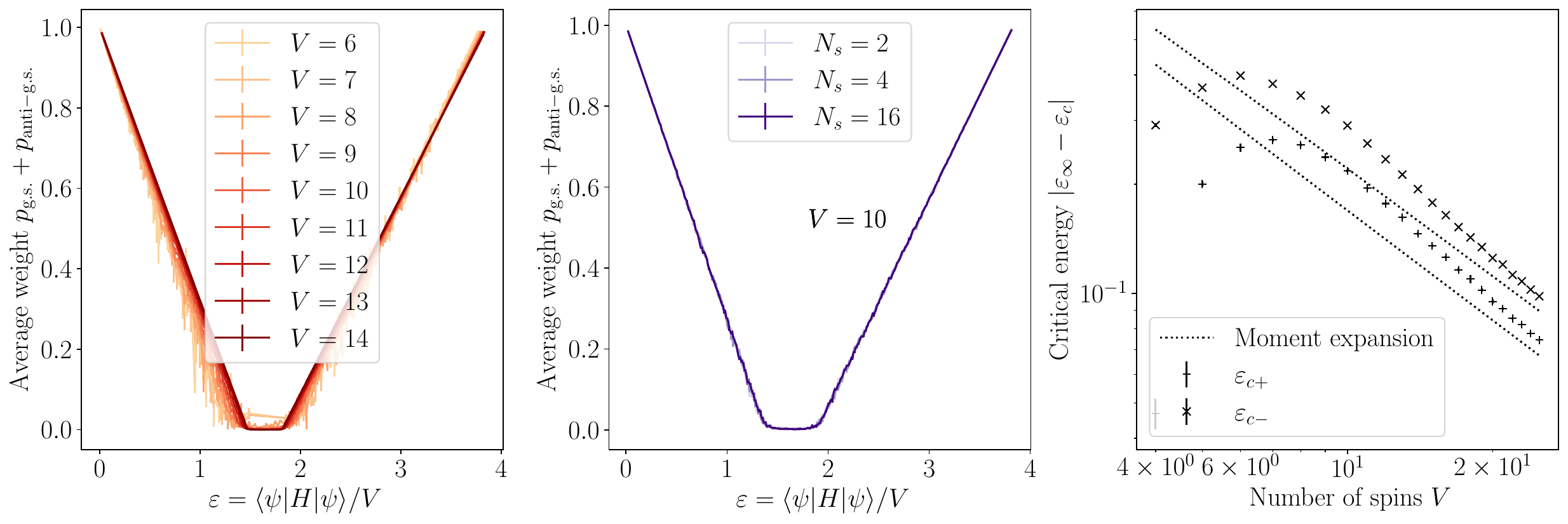}
    \caption{\textbf{One dimensional mixed-field Ising model (MFIM): sampling, convergence, and critical energy.}
    \textbf{Left}: average ground state and anti-ground-state weight as a function of $\varepsilon$ across system sizes shows the same tripartite behavior as GOE, GUE, 1D TFIM, 2D TFIM, and Heisenberg models.
    \textbf{Center}: Average weight as a function of $\varepsilon$ across the number of Galilean Monte Carlo steps per nested sampling step, $N_s$.
    Our results agree between the three $N_s$, indicating that we use a sufficient number of steps.
    \textbf{Right:} Critical energy densities $\varepsilon_{c\pm} = \frac 1 {\mathcal N_\bulk}$ as a function of system size $V$, computed via stochastic trace, together with moment expansion prediction.
    (Errorbars from nonparametric bootstrap are too small to be visible.)
    Both $\varepsilon_{c\pm}$ are approaching the moment expansion, though they do not reach it for accessible system sizes $L \le {24}$.
    }
    \label{fig:mfim}
\end{figure*}

To put the 1D transverse-field Ising model into the free-form \eqref{eq:ff}, use a Jordan-Wigner transformation.
Start by rotating $\sigma^x \leftrightarrow \sigma^z$ for
\begin{align}
    H = -J\sum_{j=1}^{V-1} \sigma^x_j \sigma^x_{j+1} - h\sum_{j = 1}^V\sigma^z_j\;.
\end{align}
(Recall that we use open boundary conditions in 1D;
this is purely for convenience,
so we do not have to consider the fermion parity factor in the $\sigma^x_1 \sigma^x_V$ term. ) 
Write Majorana fermion operators
\begin{align}
\begin{split}
    \gamma_{2j-1} &= \frac 1 {\sqrt 2}\prod_{j' < j}\left[\sigma^z_{j'}\right] \sigma^x_j\\
    \gamma_{2j} &= \frac 1 {\sqrt 2}\prod_{j' < j}\left[\sigma^z_{j'}\right] \sigma^y_j\;,
\end{split}
\end{align}
which have
\begin{align}
\begin{split}
    \gamma_l^\dagger &= \gamma_l \\
    \{\gamma_l, \gamma_m\} &= \delta_{lm}
\end{split}
\end{align}
and
\begin{align}
\begin{split}
    \sigma^z_j &= 2i\gamma_{2j-1}\gamma_{2j} \\
    \sigma^x_j \sigma^x_{j+1} &= 2i\gamma_{2j} \gamma_{2j+1}\;;
\end{split}
\end{align}
the Hamiltonian is then
\begin{align}
   H = \sum_{ij} 2iA_{ij} \gamma_i\gamma_j
\end{align}
for a real antisymmetric $A_{ij}$.
Because $A_{ij}$ is real and antisymmetric its nonzero eigenvalues come in pairs of the form $\pm ia$, and we can write
\begin{align}
2iA =
U
\begin{pmatrix}
    -\epsilon_1/2 &\\
    &\ddots\\
    && -\epsilon_V/2 \\
    &&& \epsilon_1 \\
    &&&& \ddots\\
    &&&&& \epsilon_V/2
\end{pmatrix}
U^\dagger
\end{align}
with
\begin{align}
    U_{i+V,j}^* = U_{ij}\;.
\end{align}
Write now $V$ complex fermions
\begin{align}
\begin{split}
    \eta_i &= \sum_j U_{ij}\gamma_j \\
    \eta_i^\dagger &= \sum_j U_{ij}^*\gamma_j = \eta_{i + V,j}\;,
\end{split}
\end{align}
where $1 \le i \le V$.
These have the usual fermionic anticommutation relations
\begin{align}
\begin{rcases}
    \{\eta_i, \eta_j\} = 0\\
    \{\eta_i^\dagger, \eta_j\} = \delta_{ij} \qquad
\end{rcases}
\qquad 1 \le i,j \le V
\end{align}
and up to a constant shift
\begin{align}
    H = \sum_j \epsilon_j \eta^\dagger_j \eta_j
\end{align}
as desired.

\section{Additional Hamiltonians and details of the Gaussian random matrices }\label{app:add-ham}
\subsection{Mixed-field ising model (MFIM)}
The 1D mixed-field Ising model (MFIM)
\begin{align}
    H = \sum_{j = 1}^{V-1} \sigma^z_j \sigma^z_{j+1} + \sum_{j = 1}^V [h_x\sigma^x_j + h_z \sigma^z_j]
\end{align}
is a robustly non-integrable model \cite{kimTestingWhetherAll2014};
we take the parameters
\begin{align}
\begin{split}
   h_x &= 1.4  \\
   h_z &= 0.9045
\end{split}
\end{align}
that are sometimes used to test numerical methods for thermalization and hydrodynamics \cite{rakovszkyDissipationassistedOperatorEvolution2022a,artiacoEfficientLargeScaleManyBody2024a,yi-thomasComparingNumericalMethods2024}.
Because we use open boundary conditions it has
\begin{align}
\begin{split}
    s^2 &\equiv V \mathcal N^{-1} \sum (\varepsilon_\alpha - \varepsilon_\infty)^2\\
    &= (1 - 1/V) + h_x^2 + h_z^2
\end{split}
\end{align}
(cf Eq.~\ref{eq:s2-models}).
Fig.~\ref{fig:mfim} shows $p_\gs + p_\antigs$ and $\varepsilon_{c\pm}$ for this mixed-field Ising model; its behavior is similar to the free-fermion 1D TFIM and the non-integrable 2D TFIM.

\begin{figure*}
    \centering
    \includegraphics[width=0.98\textwidth]{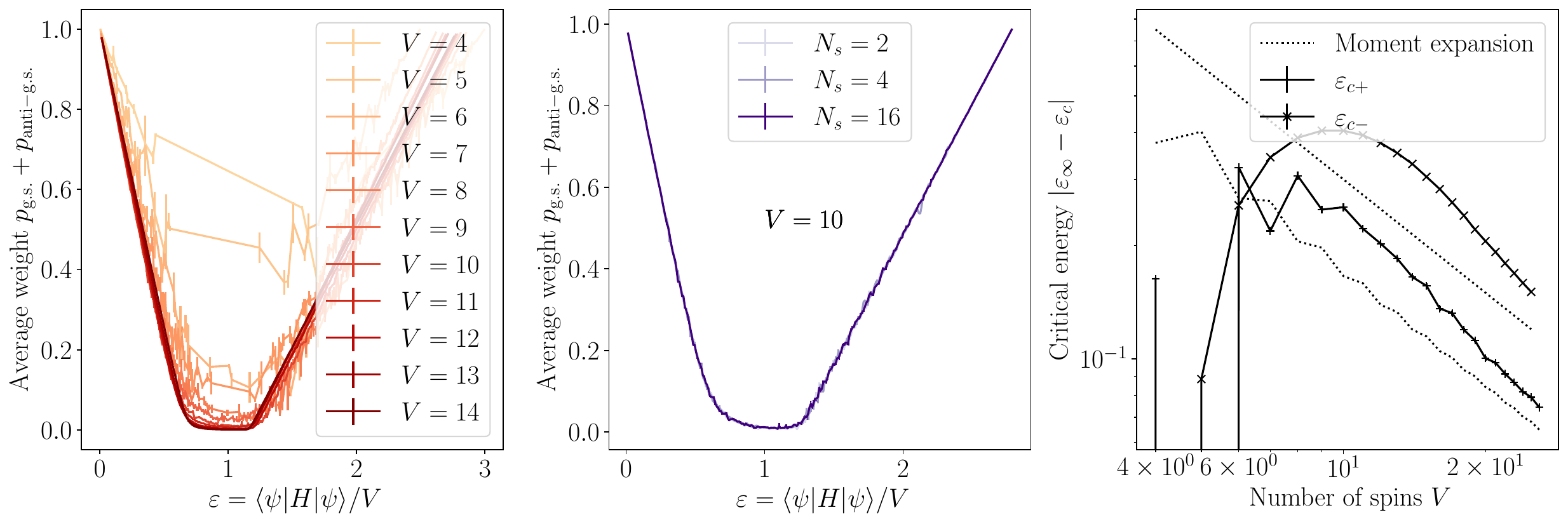}
    \caption{\textbf{One dimensional Heisenberg model: sampling, convergence, and critical energy.}
    \textbf{Left}: average ground state and anti-ground-state weight as a function of $\varepsilon$ across system sizes.
    \textbf{Center}: Average weight as a function of $\varepsilon$ across the number of Galilean Monte Carlo steps per nested sampling step, $N_s$.
    Our results agree between the three $N_s$, indicating that we use a sufficient number of steps.
    \textbf{Right:} Critical energy density $\varepsilon_{c\pm}$. For small $V$ $\varepsilon_{c\pm}$ crosses 0---that is, the critical energy is above the infinite-temperature energy---due to the large ground state degeneracy.
    $\varepsilon_{c+}$ displays a prominent even-odd alternation because it is the transition to the FM anti-ground-state or AFM ground state, whose energy and other properties display a prominent even-odd alternation.
    The approach to the moment-expansion value is slower than the MFIM (Fig.~\ref{fig:mfim})  or the TFIM (main text Fig.~3);
    we attribute this to the fact that the Heisenberg density of states approaches its limiting Gaussian form more slowly than the MFIM or TFIM (Fig.~\ref{fig:heis-dos}).
    }
    \label{fig:heisenberg}
\end{figure*}

\subsection{One dimensional Heisenberg model} \label{app:heisenberg}
The one-dimensional Heisenberg model is
\begin{align}
    H = - \sum_{j = 1}^{V-1} \bm \sigma_j \cdot \bm \sigma_{j+1} - \bm \sigma_V \cdot \bm \sigma_1\;.
\end{align}
Note that we take periodic boundary conditions and choose the sign so the model is ferromagnetic.
The model is Bethe ansatz integrable and symmetric under $SU(2)$ spin rotations, $\mathbb Z_V$ translations, and $\mathbb Z_2$ reflections.

The ground state is the ferromagnetic ground state, which is $N+1$-fold degenerate; the projector onto the degenerate ground space is
\begin{align}
\begin{split}
    P_\gs &= \sum_n \ketbra{\gs;n}{\gs;n}\;, \\
   \ket{\gs;n} &\propto (S^+_{\text{tot}})^n\ket{\downarrow \dots \downarrow} \\
   S^+_{\text{tot}} &= \sum_j S^+_j\;.
\end{split}
\end{align}
In the limit of large system sizes the model is gapless; the low-lying excitations are approximately spin waves with spectrum 
\begin{align}
\begin{split}
    \omega_k &\propto (1 - \cos k)\\
    &\approx \frac 1 2 k^2\;, k \ll 1\;.
\end{split}
\end{align}
For finite sizes, therefore, the ground space is separated from the low-lying excited states by a gap that shrinks polynomially in system size.

The single-excitation density of states has a van Hove singularity at $\omega = 0$.
One might worry that the resulting divergence in quantities of the form $\int dE\; \rho(E) / E$ 
breaks our arguments about finite size scaling both of $E_c$ and of $p_\gs$ near $E_c$.
In fact (as we argue in App.~\ref{app:gapless-hamiltonians}) this singularity diverges gently with system size. 
So although its contributions do give subleading corrections to the finite-size scaling, they do not change our conclusions.

The anti-ground-state is the ground state of the antiferromagnetic Heisenberg model \cite{auerbachInteractingElectronsQuantum1998}.
For $V$ even it is non-degenerate and has total $\bm S^2 = 0$, by Marshall's theorem \cite{marshallAntiferromagnetism1955,liebOrderingEnergyLevels1962}, hence has total $S^z = 0$;
by Lieb-Schultz-Mattis it has an excited state with vanishing gap \cite{liebTwoSolubleModels1961},
and indeed an exact Bethe ansatz solution gives that magnons are gapless with spectrum \cite{descloizeauxSpinWaveSpectrumAntiferromagnetic1962}
\begin{align}
    \omega_k = \frac \pi 2 |\sin k|\;.
\end{align}
For $V$ odd the anti-ground-state is fourfold degenerate due to $\mathds Z_2$ reflection symmetry and the $SU(2)$ symmetry (which includes elements that take $S^z \leftrightarrow - S^z$).

Because the ground state and anti-ground-state are degenerate, we consider the total weight on the (anti)-ground space:
\begin{align}
\begin{split}
    p_\gs &= \braket{\psi |P_\gs | \psi} \\
    p_\antigs &= \braket{\psi |P_\antigs | \psi} \;.
\end{split}
\end{align}
(We briefly discuss the logic behind this choice in the main text, in the context of the nearly-doubly-degenerate ferromagnetic TFIM.)
Likewise we take $\mathcal N_\bulk$ and sums $\sum_{\alpha \ne \gs, \antigs}$ to exclude the whole (anti)-ground space.

Fig.~\ref{fig:heisenberg} shows $p_\gs + p_\antigs$ and $\varepsilon_{c\pm}$ for the 1D Heisenberg model.
Its behavior is qualitatively similar to the free-fermion 1D TFIM and the non-integrable 2D TFIM.

There is, however, a distinct asymmetry between the ground state transition and the anti-ground-state transition (Fig.~\ref{fig:heisenberg} left, center):
the ground state transition is appreciably rounded out.
We attribute this to the different densities of states---in particular, the fact that the single-excitation spectrum is quadratic near $E_\gs$ but linear near $E_\antigs$, 
so there is a van Hove singularity at $E = E_\gs$ but not at $E = E_\antigs$.
Consequently as the average energy goes down weight is concentrated on ever-smaller subspaces, all nearly degenerate with the ground state.
At finite system size, then, the transition is rounded out for the ground state to a greater degree than for the anti-ground state.

\begin{figure*}
    \centering
    \includegraphics[width=0.75\textwidth]{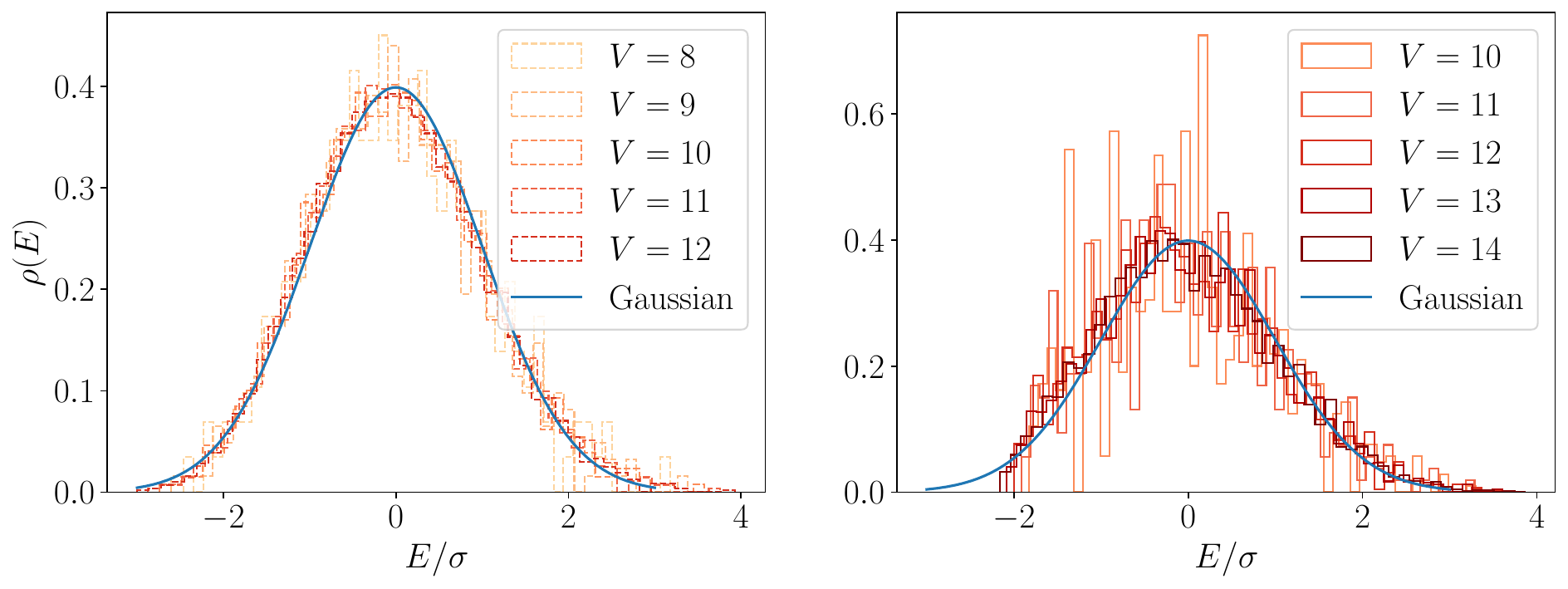}
    \caption{\textbf{Density of states} for the 1D Heisenberg and 1D mixed-field Ising (MFIM) models. In each case the energies are rescaled by the standard deviation $\sigma $ of the density of states.
    The Heisenberg only slowly approaches its limiting Gaussian form.
    }
    \label{fig:heis-dos}
\end{figure*}

The critical energy densities of the Heisenberg model (Fig.~\ref{fig:heisenberg} right) do not approach the leading-order moment-expansion value as quickly as the TFIM (main text Fig.~3) or the MFIM (Fig.~\ref{fig:mfim}).
In our view these numerics are consistent with the moment expansion but do not confirm it---for the Heisenberg model the analytical calculations of App.~\ref{app:fe-mu} and \ref{app:Ec-system-size} must stand on their own.
We believe $\varepsilon_{c\pm}$ approaches its limiting moment-expansion behavior slowly because the density of states approaches its limiting Gaussian form slowly,
and that slow approach in turn results from the model's Bethe ansatz integrability.
Fig.~\ref{fig:heis-dos} shows the density of states of the Heisenberg model across system size, compared with the MFIM and the limiting Gaussian%
---in the MFIM the approach to the limiting Gaussian is clear, but in the Heisenberg model it is not.

\subsection{Gaussian ensembles of random matrices}\label{app:GbE}

In addition to spin Hamiltonians we consider random matrices drawn from the Gaussian orthogonal and unitary ensembles (GOE and GUE).
These are
\begin{align}
    H = \frac V {2^{3/2}\sqrt {\mathcal N}} \sum_{i,j = 1}^{\mathcal N} (X_{ij} + X_{ji}^*) \ketbra{i}{j}
\end{align}
where once again $\mathcal N = 2^V$ is the Hilbert space dimension and the $X_{ij}$ are i.i.d. Gaussian:
\begin{align}
   X_{ij} \sim
   \begin{cases}
    \mathcal N(0,1;\mathbb R) & \text{GOE} \\
    \mathcal N(0,1;\mathbb C) & \text{GUE}\;.
   \end{cases}
\end{align}
Note that we choose a slightly unorthodox normalization convention, so that the bulk density of states is bounded $-V \le E \le V$ (setting aside outliers)---this simplifies a number of later expressions.
Many quantities are conveniently framed in terms of 
\begin{align}
    \beta =
   \begin{cases}
    1 & \text{GOE} \\
    2 & \text{GUE}\;;
   \end{cases}
\end{align}
it is convenient to refer to the ensembles collectively as ``G$\beta$E''.
The G$\beta$E ensembles are unitarily invariant.
In each case, the large-system density of states is
\begin{align}
    \rho(E) = \frac 2 {\pi V^2} \sqrt{V^2 - E^2}\;;
\end{align}
this is the famous Wigner semicircle law \cite{mehta2004,tao2012}
In this large-system limit { outliers approach the bulk spectrum} and
the spectrum is $-V \le E \le V$ before shifting the ground state energy to 0,
or $0 \le E \le 2V$ after.
(For finite systems this simple picture is complicated by outlier eigenvalues, on which see below.)

\begin{figure*}[t]
    \centering
    \includegraphics[width=0.75\textwidth]{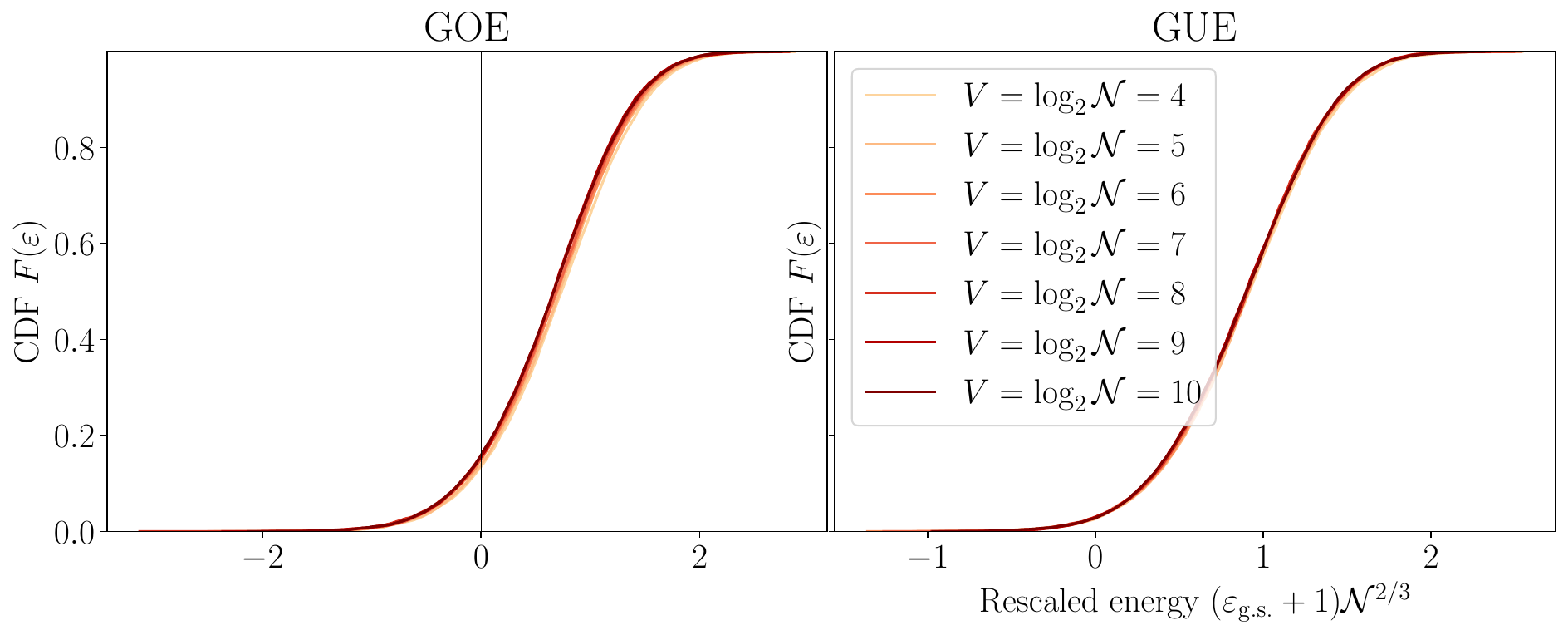}
    \caption{\textbf{CDFs of ground state energy density} for GOE (\textbf{left}) and GUE (\textbf{right}) matrices display Tracy-Widom scaling. }
    \label{fig:tracy-widom}
\end{figure*}

Each \GbE{} ensemble has the same spectral properties as a particular real tridiagonal matrix ensemble;
one can see this by Householder tridiagonalization of the \GbE{} matrices
\cite{trotterEigenvalueDistributionsLarge1984,dumitriuMatrixModelsBeta2002}.
The result is
\begin{align}
   H' = \frac V {2\sqrt{\mathcal N \beta}}
   \begin{bmatrix}
   a_{\mathcal N}  & b_{\mathcal N-1} &  &  & \\
   b_{\mathcal N - 1} & a_{\mathcal N-1}  & b_{\mathcal N-2} &  &  \\
    & b_{\mathcal N-2} & \ddots & \ddots&  \\
    &  & \ddots & \ddots & b_1 \\
    &  &  & b_1 & a_1 
   \end{bmatrix}
\end{align}
where the diagonals $a_j$ are Gaussian and the subdiagonals $b_j$ are chi-distributed
\begin{align}
\begin{split}
    a_j &\sim \mathcal N(0,2)\\
    b_j &\sim \mathcal \chi_{\beta j}\;;
\end{split}
\end{align}
this form is useful for numerics.

\begin{figure}[t]
    \centering
    \includegraphics[width=0.98\columnwidth]{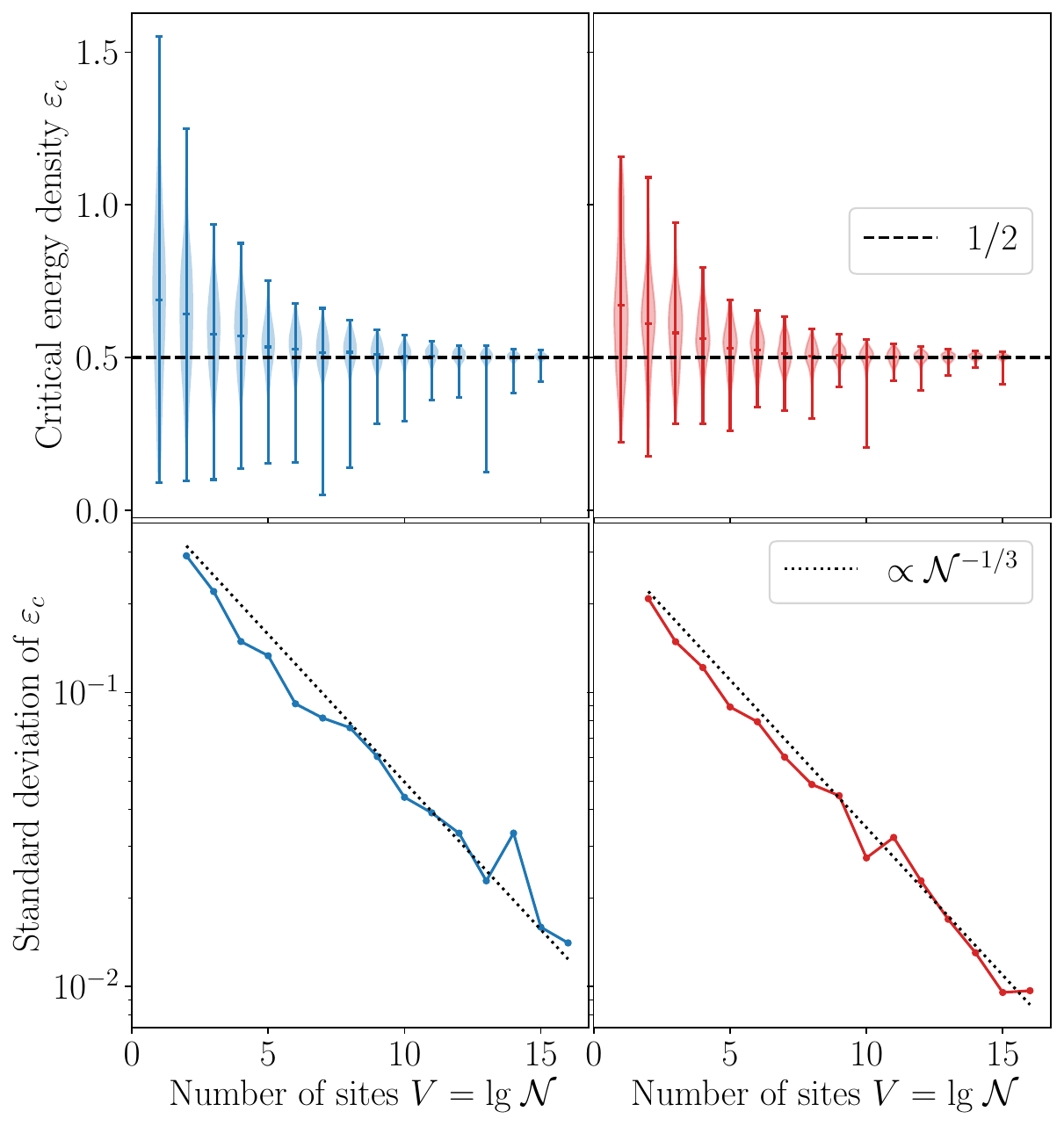}
    \caption{\textbf{Variability of critical energy density $\varepsilon_c$ across {200} \GbE{} matrices}. Top row: distribution of $\varepsilon_c$ as a function of number of sites. Bottom row: standard deviation of $\varepsilon_c$ is empirically $\propto \mathcal N^{1/3}$.
    In each case the critical energy is computed with full exact diagonalization;
    this is possible at sizes $V = {16}$ spins
    (Hilbert space dimension $\mathcal N = {65536}$)
    because the matrices are already tridiagonal,
    so the most expensive part of full diagonalization, the Householder tridiagonalization, has already been done.)
    }
    \label{fig:Ec-GbE-violinplot}
\end{figure}

The ground state energy of a G$\beta$E matrix is drawn from the Tracy-Widom distribution
\cite{tracyLevelspacingDistributionsAiry1993a,tracyLevelSpacingDistributionsAiry1994,tracyOrthogonalSymplecticMatrix1996,tracyDistributionsRandomMatrix2009}.
In the limit of large Hilbert space, the CDF of the ground state energy density $F_{N,\beta}(\varepsilon_\gs)$ has the scaling form
\begin{align}
    F_\beta(x) = \lim_{\mathcal N \to \infty} F_{N,\beta}\left( -1 - \frac {x}{\mathcal N^{2/3} \sqrt {2}}\right)
\end{align}
(cf Fig.~\ref{fig:tracy-widom}; recall that we have not yet shifted the ground state energy to $E_\gs = 0$).
Heuristically, therefore, the ground state approaches the edge of the semicircle like
\begin{align}
    \varepsilon_\gs = -1 + O(\mathcal N^{2/3}).
\end{align}

\begin{figure}[t!]
    \centering
    \includegraphics[width=0.98\columnwidth]{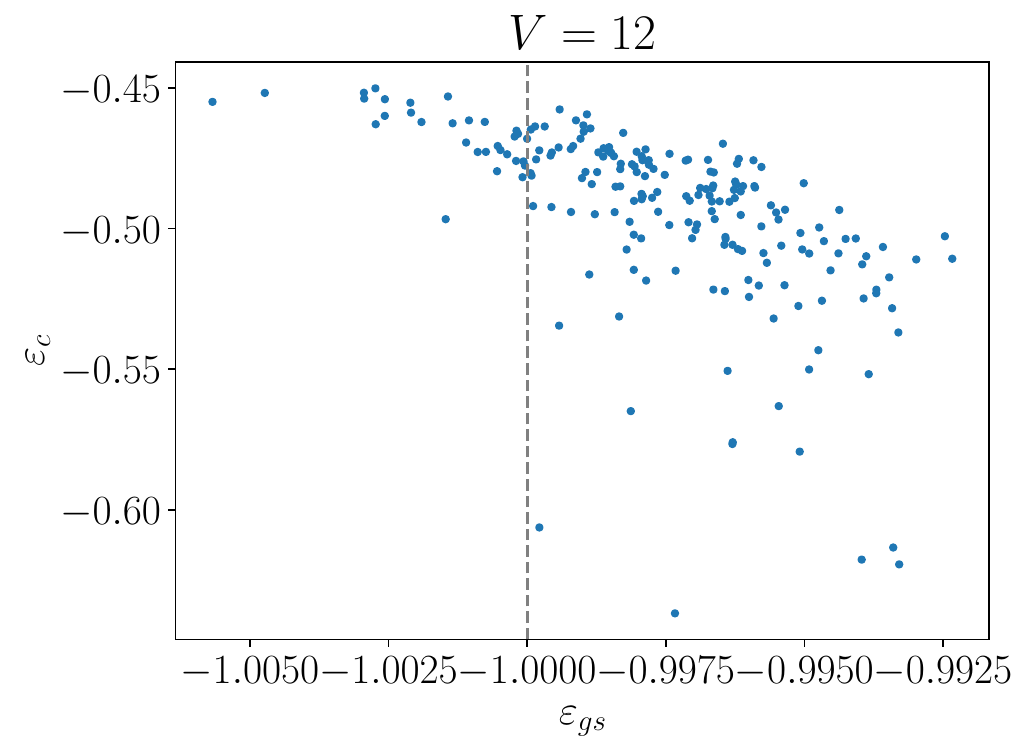}
    \caption{\textbf{$\varepsilon_c$ and $\varepsilon_\gs$ for GOE random matrices} on $V = {12}$ spins. Both are in unshifted units. Dashed line marks the edge of the bulk spectrum.
    }
    \label{fig:Ec-Egs-GOE}
\end{figure}

In the large-system limit the critical energy density is
\begin{align}
    \varepsilon_c = \left[\frac 2 \pi \int d\varepsilon \; {\sqrt{1 - (\varepsilon - 1)^2} \times \frac 1 \varepsilon } \right]^{-1} = \frac 1 2
\end{align}
(after shifting the ground state energy density to 0).
For finite systems the critical energy density
\begin{align}
\varepsilon_c = \left[\sum_{\alpha \ne g.s.} \frac 1 {\varepsilon_\alpha}\right]^{-1}
\end{align}
displays considerable variation (Fig.~\ref{fig:Ec-GbE-violinplot}).
Empirically we find 
\begin{align}
\text{std}\; \varepsilon_c \propto \mathcal N^{-1/3}
\end{align}
(where the standard deviation is over \GbE{} matrices).
The variation appears to be related to, but not solely driven by, the Tracy-Widom variation of the ground state energy (cf Fig.~\ref{fig:Ec-Egs-GOE}).
We attribute it to variation in the whole low-energy subspace,
but we have not attempted to estimate it analytically.

\begin{figure*}
    \centering
    \includegraphics[width=0.75\textwidth]{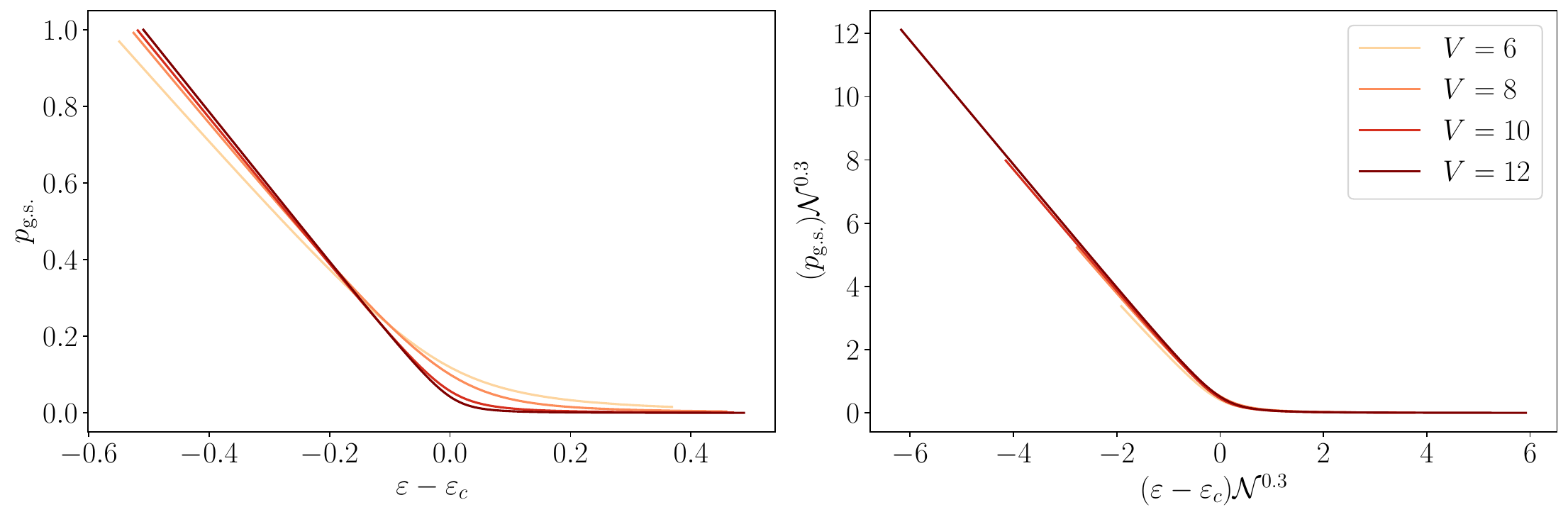}
    \caption{\textbf{Critical behavior of typical GOE matrices.} \textbf{Left}: ground state probability as a function of energy. \textbf{Right:} empirical scaling collapse.
    }
    \label{fig:GOE-collapse}
\end{figure*}

Fig.~\ref{fig:GOE-collapse} shows semianalytical average ground state weights as a function of average energy, for typical GOE Hamiltonians.
To compute these weights we first diagonalize {101} tridiagonal GOE Hamiltonians.
These have a wide range of $E_c$ (vide supra) and presumably other behavior;
we choose the Hamiltonian with the median $E_c$, since that Hamiltonian presumably displays ``typical'' behavior, in some sense.
Given that Hamiltonian's eigenenergies $E_\alpha$ we then compute
\begin{align}
\begin{split}
Z(\beta) &= \sum_\alpha \frac 1 {1 + \beta E_\alpha} \\
E_\av(\beta) &= \frac 1 {Z(\beta)} \sum_\alpha \frac {E_\alpha} {1 + \beta E_\alpha} \\
p_\gs(\beta) &= \frac 1 {Z(\beta)}
\end{split}
\end{align}
as functions of $\beta$, and plot $p_\gs$ against $E_\av(\beta)$.
These ground state weights show an empirical scaling collapse
\begin{align}
p_\gs(E_\av) \approx \mathcal N^{-1/4} f\left([E - E_c] \mathcal N^{1/4}\right)
\end{align}
This scaling is different from the $\sqrt{\mathcal N}$ of local spin systems.

\section{Details of weight derivation and inverse temperature}\label{app:weight-deriv}

Consider a large number $\mathcal M$ of pure states $\ket {\psi^{(k)}}$.
Let us estimate the average over this ensemble of the weight on energy eigenstate $\alpha$
\begin{align}
    p_\alpha = \frac 1 {\mathcal M}\sum_{k=1}^{\mathcal M} |\braket{\alpha|\psi^{(k)}}|^2\;.
\end{align}
Dropping the normalization constraint---we will introduce it via a Lagrange multiplier---%
the set of whole-ensemble states with a given value of $p_i$ is a sphere of radius $\sqrt{p_i}$ in $2\mathcal M$-dimensional space (2 because the wavefunctions are complex).
Since $\mathcal M$ is large,
\begin{equation}
    \log \Omega \approx 2\mathcal M \sum_i\log\sqrt{p_i}=\mathcal N\sum_i \log p_i.
\end{equation}
The typical $p_\alpha$, then, is given by
\begin{align}
\begin{split}
    \argmax_{p_\alpha} &\left(\mathcal M \log p_\alpha  + \lambda' \left[1 - \sum_k p_\alpha\right] \right. \\
    &\qquad\qquad\quad\left. + \beta'\left[E - \sum_\alpha E_\alpha p_\alpha\right]\right)\;,
\end{split}
\end{align}
where $\lambda'$ and $\beta'$ the Lagrange multipliers for normalization and energy respectively.
Carrying out the maximization gives
\begin{align}
    p_\alpha = \frac {\mathcal M}{\lambda' + \beta' E_\alpha}\;;
\end{align}
relabeling the parameters gives the form in the main text.

As defined here $p_\alpha$ is not the weight of a single state on eigenstate $\alpha$,
nor is it even the average weight for energy-constrained random states.
Rather it is the most likely ensemble-average weight,
where ``most likely'' refers to the probability distribution over ensembles of $\mathcal M$ states. 
Relating this to an individual energy-constrained random state requires a typicality argument:
since $\mathcal M$ is large, most such ensembles have nearly the same $p_\alpha$, and we are justified in taking $p_\alpha$ to be the average over all energy-constrained random states.
Note also, though, that our calculation constrains the ensemble average normalization and energy expectation value,
rather than the normalization and energy expectation value of each state.

For a substantially more careful derivation, which avoids all of these issues at the cost of some complexity, see \cite{fineTypicalStateIsolated2009}.

\section{Monotonicity of and lower bound on bulk contribution to $E(\beta)$} \label{app:Ebulk-bound}

In the main text we remark without proof that when $p_\gs$ is negligible the energy is given by the bulk contribution
\begin{gather}
\begin{split}
    E_\av \approx Z_\bulk(\beta)^{-1} \sum_{\alpha \ne \gs} \frac{E_\alpha}{1 + \beta E_\alpha} \\
    Z_\bulk(\beta) = \sum_{\alpha \ne \gs} \frac 1 {1 + \beta E_\alpha}
\end{split}
\end{gather}
which is lower bounded by $\lim_{\beta \to \infty} \beta Z_\bulk(\beta)^{-1}$.
Here we prove the lower bound.

The first step in proving the lower bound is to show that the bulk contribution to the energy
\begin{align}
   E_\bulk = Z_\bulk(\beta)^{-1}  \sum_{\alpha \ne \gs} \frac{E_\alpha}{1 + \beta E_\alpha} 
\end{align}
is monotonically decreasing in $\beta$.
An elementary calculation gives
\begin{align}
    \partial_\beta E_\bulk
    &= \frac 1 {Z_\bulk^2} \sum_{\alpha, \alpha' \ne \gs}
    \frac 1 {(1 + \beta E_\alpha)(1 + \beta E_\alpha')^2}  \notag \\
    &\qquad\qquad\qquad\qquad\times E_{\alpha'} (E_\alpha - E_\alpha') \\
    &= \frac 1 {Z_\bulk^2} \sum_{\alpha, \alpha' \ne \gs}
    \frac 1 {(1 + \beta E_\alpha)^2(1 + \beta E_\alpha')^2} \notag\\
    &\qquad\qquad\qquad\qquad\times E_{\alpha'} (1 + \beta E_\alpha)(E_\alpha - E_\alpha')\;. \notag
\end{align}
Now exploit the symmetry
\begin{align}
    \sum_{\alpha\alpha'} f(\alpha, \alpha') = \frac 1 2 \sum_{\alpha\alpha'} [f(\alpha, \alpha') + \alpha \leftrightarrow \alpha']\;.
\end{align}
The first part of the summand, $\frac 1 {(1 + \beta E_\alpha)^2(1 + \beta E_\alpha')^2}$, is symmetric;
the second part becomes
\begin{align}
\begin{split}
    &E_{\alpha'} (1 + \beta E_\alpha)(E_\alpha - E_\alpha') \  + \  [\alpha \leftrightarrow \alpha'] \\
    &\quad = - (E_{\alpha'} - E_\alpha)^2
\end{split}
\end{align}
so
\begin{align}
\begin{split}
    \partial_\beta E_\bulk
    &= - \frac 1 {Z_\bulk^2} \sum_{\alpha, \alpha' \ne \gs}
    \frac {(E_\alpha - E_{\alpha'})^2} {(1 + \beta E_\alpha)^2(1 + \beta E_\alpha')^2} \\
    &< 0\;.
\end{split}
\end{align}
It is useful also to note that
\begin{gather}
\begin{split}
    \lim_{\beta  \to \infty} \beta Z_\bulk(\beta) = \sum_{\alpha \ne \gs} \frac 1 {E_\alpha}\\
    \lim_{\beta  \to \infty} \beta \sum_{\alpha \ne \gs} \frac {E_\alpha}{1 + \beta E_\alpha} = \mathcal N_\bulk
\end{split}
\end{gather}
where $\mathcal N_\bulk$ is the ground state degeneracy.
Having shown these things, we immediately see
\begin{align}
\begin{split}
    E_\bulk &\ge \lim_{\beta \to \infty} Z_\bulk(\beta)^{-1} \sum_{\alpha \ne \gs} \frac {E_\alpha}{1 + \beta E_\alpha}  \\
    &= \left[\frac 1 {\mathcal N_\bulk} \sum_{\alpha \ne \gs} \frac 1 {E_{\alpha}}\right]^{-1}\;.
\end{split}
\end{align}

\section{Free energy and scaling of spectral moments}\label{app:fe-mu}

Here and elsewhere we use that---for many spin systems---the central moments of the energy density
\begin{align}
\begin{split}
    \mu_n(\varepsilon)
    &= \frac 1 {\mathcal N} \sum_\alpha (\varepsilon_\alpha - \varepsilon_\infty)^n \\
    &= \frac 1 {V^n} \frac 1 {\mathcal N} \sum_\alpha (E_\alpha - E_\infty)^n
\end{split}
\end{align}
are small
\begin{align}
    \mu_n(\varepsilon) \sim V^{-n/2}\;.
\end{align}
for even moments, and smaller for odd moments.
(Here once again $E_\infty = \mathcal N^{-1} \sum_\alpha E_\alpha$ and $\varepsilon_\infty = E_\infty / V$.)

\subsection{Second moment and spin systems}
One can see the scaling explicitly for the second moment of the density of states of a local spin model with reasonable parameters.
Write such a model
\begin{align} \label{eq:spin-halpha}
    H = \sum_\alpha h_\alpha \sigma^\alpha
\end{align}
where the $\sigma^\alpha$ are distinct nontrivial Pauli strings.
(Note that---for the moment---we are not shifting the ground state to $E = 0$.)
Since \( \tr \sigma^\alpha = 0\) and \(\tr \sigma^\alpha \sigma^\beta = 2^V \delta_{\alpha\beta}\)\;,
we immediately have
\begin{align}\label{eq:trH2-pauli-coeffs}
\begin{split}
    \tr H &= 0\\
    2^{-V} \tr H^2 &= \sum_\alpha h_\alpha^2\;.
\end{split}
\end{align}
Since the model is local there are $O(V)$ different $h_\alpha$.
Provided the $h_\alpha$ themselves have finite second moment%
---e.g. they are bounded---%
then
\begin{align}
    2^{-V} \tr H^2 &\sim V \\
    \mu_2(\varepsilon) = \frac 1 {V^2} 2^{-V} \tr H^2 &\sim 1/V.
\end{align}
Here as elsewhere we define the (asymptotically) $V$-independent quantity
\begin{align}
    s^2 = V \mu_2(\varepsilon) \;;
\end{align}
for spin systems of the form \eqref{eq:spin-halpha} this is
\begin{align}\label{eq:s2-pauli-coeffs}
    s_2 = \frac 1 V \sum_\alpha h_\alpha^2\;.
\end{align}

\subsection{Scaling of spectral moments in general}
To see the moment scaling in general, assume that the free energy
\begin{align}
    F(\beta) = \ln \tr e^{-\beta H}
\end{align}
is extensive
\begin{align}
    F(\beta) = V \ln 2 + V f(\beta)\;,
\end{align}
where $f(\beta)$ is asymptotically constant in $V$.
{(Note that we have separated out a Hilbert space factor $\ln \mathcal N = V \ln 2$.)}
Here, unlike most of the rest of the paper, let us take $0 = \tr H$
so the moment generating function directly gives the central moments we seek.

$Vf(\beta)$ is the cumulant generating function of the density of states.
It has (in the large-$V$ limit) a Taylor series
\begin{align}
f(\beta) =  \frac 1 2 f_2\beta^2 + \frac 1 {3!} f_3\beta^3 + \dots\;.
\end{align}
(The first two terms disappear: $f(\beta = 0)$ because we have taken the normalization outside $f$,
and $\partial_\beta f(\beta)|_{\beta = 0}$ because we have taken the mean energy to be $0 = \tr H$.)

{ To estimate moments of the density of states we use the moment generating function.
Recall that the moment generating function of a random variable $X$ is
\begin{align}
\begin{split}
M_X(\beta) &\equiv \mathds E_X[e^{\beta X}] \\
&= \sum \frac {\beta^n} {n!} \mu_n(X)
\end{split}
\end{align}
where $\mathds E_X$ is the expectation value over $X$ and $\mu_n(X) = \mathds E_X(X^n)$ is its $n$th moment;
it transforms under scaling as
\begin{align}
    M_{aX}(\beta) = M_X(a\beta)\;.
\end{align}
If we treat the energy as a random variable distributed according to the density of states,
its moment generating function is
}
\begin{align}
\begin{split}
    M_E(\beta) &= { 2^{-V} \sum_\alpha e^{\beta E_\alpha} }\\
    &=e^{-Vf(\beta)}\;.
\end{split}
\end{align}
{Using the scaling behavior of the moment generating function,}
the moment generating function of the energy density is
\begin{align}
\begin{split}
    M_\varepsilon(\beta)
    &= e^{-Vf(V^{-1}\beta)}\\
    &= \exp\left[- \frac 1 {2} V^{-1}f_2 \beta^2\right] \left(1 + O(V^{-2})\right)\;.
\end{split}
\end{align}
{The first factor is the moment generating function of a Gaussian,}
so the moments of the density of states are---up to $O(V^{-2})$ corrections---%
the moments of a Gaussian with variance $V^{-1} f_2$.
Consequently the central moments of the energy density are asymptotically
\begin{align}
    \mu_n(\varepsilon) \approx \begin{cases}
         (f_2/V)^{n/2} (n-1)!! & \text{$n$ even} \\
         0 &\text{$n$ odd}\;,
    \end{cases}
\end{align}
as desired.

\section{Drift of critical energy as a function of system size}\label{app:Ec-system-size}

The critical energy density is 
\begin{align}
    \varepsilon_{c-}^{-1} = \sum_{\alpha \ne \gs} \frac 1 {\varepsilon_\alpha}\;.
\end{align}
Assume the ground state is nondegenerate and write
\begin{align}
    \varepsilon_\infty' = \frac 1 {\mathcal N - 1}\sum_{\alpha \ne g.s.} \varepsilon_\alpha\;.
\end{align}
(Note that $\varepsilon_\infty'$ differs from $\varepsilon_\infty = (V\mathcal N)^{-1} \tr H = \frac 1 {\mathcal N} \sum_{\alpha} \varepsilon_\alpha$ by $O(\mathcal N^{-1})$,
because the average in $\varepsilon_\infty'$ does not include the ground state.)
Expand the critical energy density around this $\varepsilon_\infty'$:
\begin{align}
\begin{split}
    \varepsilon_{c-}^{-1}
    &= \frac 1 {\varepsilon'_\infty} \sum_{\alpha \ne \gs} \frac 1 {1 -(\varepsilon'_\infty - \varepsilon_\alpha)/\varepsilon'_\infty}\\
    &= \frac 1 {\varepsilon'_\infty} \sum_{\alpha \ne \gs} \sum_n (\varepsilon'_\infty)^{-n} (\varepsilon'_\infty - \varepsilon_\alpha) \\
    &= \frac 1 {\varepsilon'_\infty} \sum_n (-\varepsilon'_\infty)^{-n} \mu'_n(\varepsilon)\;,
\end{split}
\end{align}
where 
\begin{align}
    \mu_n'(\varepsilon) = \frac 1 {\mathcal N - 1}\sum_{\alpha \ne\gs} (\varepsilon_\alpha' - \varepsilon'_\infty)^n
\end{align}
are the central moments of the bulk density of states.
Since those central moments fall off as $O(V^{-n/2})$ the leading correction is
\begin{align}
    \varepsilon_{c-}^{-1}
    &= \frac 1 {\varepsilon_\infty'} \left( 1 + \mu_2'(\varepsilon) / \varepsilon_\infty'^2\right) + O(V^{-3/2})\\
    &\approx \frac 1 {\varepsilon_\infty} \left( 1 + \frac {s^2}{V\varepsilon_\infty} \right) + O(\mathcal N^{-1})
\end{align}
where (as elsewhere in the paper)
\begin{subequations}
\begin{align}
    \varepsilon_\infty &= \mathcal N^{-1} \tr H/V\\
    s^2 &= V\mathcal N^{-1} \tr\left[ (H/V - \varepsilon_\infty)^2\right]\;.
\end{align}
\end{subequations}
(We have neglected an $O(\mathcal N^{-1})$ correction coming from the difference between the moments $\mu'$ of the bulk density of states and $\mu$ of the whole density of states, including ground states).
Using Eq.~\eqref{eq:s2-pauli-coeffs}, the three Hamiltonians we consider in numerics (paramagnet, TFIM, and Heisenberg) have
\begin{subequations}\label{eq:s2-models}
\begin{align} 
    s^2_\para &= h^2 \\
    s^2_{\TFIM, \mathrm{MFIM}} &= \frac{V-1}{V} + h_x^2 + h_z^2 \\
    s^2_\Heis &= 3\;.
\end{align}
\end{subequations}
where the factor $\frac {V-1}{V}$ in the TFIM / MFIM value comes from the fact that with open boundary conditions there are only $V-1$ bond terms, but $V$ site (field) terms.

Proceeding analogously, the upper critical energy density is
\begin{align}
\begin{split}
    \varepsilon_{c+} &= \frac 1 {\mathcal N-1} \sum_{\alpha \ne \antigs} \frac 1 {\varepsilon_\antigs - \varepsilon_\alpha} \\
    &\approx E_\infty  \left (1 - \frac {s^2}{V(\varepsilon_\infty - \varepsilon_\antigs)^2}\right)
\end{split}
\end{align}
---that is, both the ground state and anti-ground-state phases approach the middle of the spectrum like $1/V$.

\section{Semianalytical calculation of $p_\gs$ and $E_\av$ as functions of $\beta$ for the TFIM}\label{app:pgs-Eav-tfim}
The TFIM is free-fermion-integrable
\begin{align}
    H = \sum \epsilon_k n_k
\end{align}
via the Jordan-Wigner transformation (cf App.~\ref{app:Ec-ff}).
We seek
\begin{align}
   Z(\beta) &= \sum_\alpha \frac 1 {1 + \beta E_\alpha} \\
   Z(\beta)E(\beta) &= \sum_\alpha \frac {E_\alpha} {1 + \beta E_\alpha} \\
   p_\gs(\beta) &= \frac 1 {Z(\beta)}\;.
\end{align}
As in App.~\ref{app:Ec-ff}, we can use a resolvent trick and numerical integration with QuadGK.jl \cite{JuliaMathQuadGKjl2025} to evaluate $Z$ and $ZE$:
\begin{widetext}
\begin{subequations}
\begin{align}
    Z(\beta) &= \int_0^\infty dx\; \tr e^{-x(1 + \beta H)}  = \int_0^\infty dx\; e^{-x} \prod_k \left[1 + e^{-x\beta \epsilon_k} \right] \\
    ZE(\beta) &= \int_0^\infty dx\; \tr He^{-x(1 + \beta H)}  = \int_0^\infty dx\;e^{-x} \sum_k  \epsilon_k e^{-x\beta \epsilon_k} \prod_{k \ne k'} \left[1 + e^{-x\beta \epsilon_k'} \right]\;.
\end{align}
\end{subequations}
Because we require $\beta \gg \mathcal N = 2^V$ for large volumes $V \sim 50-100$,
however, we must take care to evaluate these expressions in a way that is not unduly influenced by floating-point roundoff error.
We therefore use the equivalent expressions
\begin{subequations}
\begin{align}
    Z &= 1 + \frac 1 \beta \int dy\; e^{-y/\beta} \text{expm1}\left(\sum_k\text{log1pexp}(-y\epsilon_k)\right) \\
    ZE &= \sum_k \epsilon_k\left[\frac {1} {1 + \beta \epsilon_k} + \frac{1}{\beta} \int dy\; e^{-y/\beta}\text{expm1}\left(\sum_{k' \ne k} \text{log1pexp}(-y\epsilon_{k'})\right)\right]
\end{align}
\end{subequations}
\end{widetext}
where 
\begin{align}
\begin{split}
    \text{expm1}(w) &= e^w - 1\\
    \text{log1pexp}(w) &= \ln(1 + e^w)
\end{split}
\end{align}
refer to numerically stable implementations of the two functions, the first from the Julia standard library and the second from LogExpFunctions.jl \cite{JuliaStatsLogExpFunctionsjl2025}.

\section{Finite-size scaling near the transition}\label{app:scaling-collapse}

\subsection{Analytical calculation via moment expansion}
In this section we use a moment expansion to calculate $p_\gs$ as a function of $E_\av$ near the eigenstate condensation transition.

Note that here, unlike the rest of the paper, we nondimensionalize energy by $E_\infty = \mathcal N^{-1} \tr H$
---that is, we take $E \to E /E_\infty$.
{(This makes the calculations appreciably less opaque.)}
Since $E_\infty$ is extensive, the moments of $E$ in the density of states now scale like the moments of energy density (App.~\ref{app:fe-mu}).
We also re-parametrize the typical eigenstate weight
\begin{align}
    p_\alpha = \frac 1 {\lambda + \beta E_\alpha}\;;
\end{align}
consequently $\beta$ here is different from $\beta$ elsewhere.

Since $E_\gs = 0$, the extra parameter $\lambda$ is in fact
\begin{align}
    \lambda = \frac 1 {p_\gs}.
\end{align}
First use the normalization to find $\beta$ as a function of $p_\gs$:
\begin{align}
\begin{split}\label{eq:app-norm-pgs-beta}
    1 - p_\gs &=  \sum_{\alpha \ne gs} \frac 1 {\lambda + \beta E_\alpha} \\
    &= \frac {\mathcal N_\bulk}{\lambda + \beta} + O(\mu_2)
\end{split}
\end{align}
expanding in $E_\alpha - E_\infty$ and using that $E_\infty = 1$.
Here $\mu_2$ is the second central moment of $E$ in the bulk density of states.
Recall that we have non-dimensionalized by $E_\infty$, which is extensive, 
so $\mu_2$ has the scaling of an energy density moment.
Rearranging gives
\begin{align}\label{eq:pgs-beta}
    \beta = \frac {p_\gs (\mathcal N_\bulk + 1) - 1} {p_\gs(1-p_\gs)}\;.
\end{align}

\begin{figure*}
    \centering
    \includegraphics[width=0.45\textwidth]{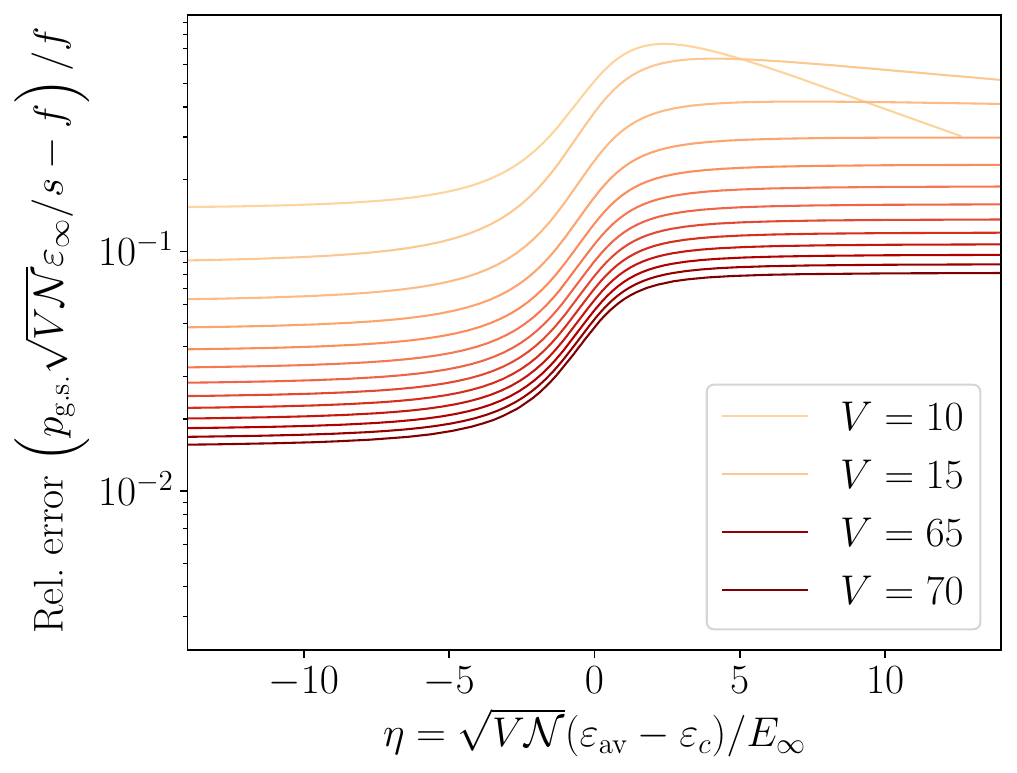}
    \includegraphics[width=0.45\textwidth]{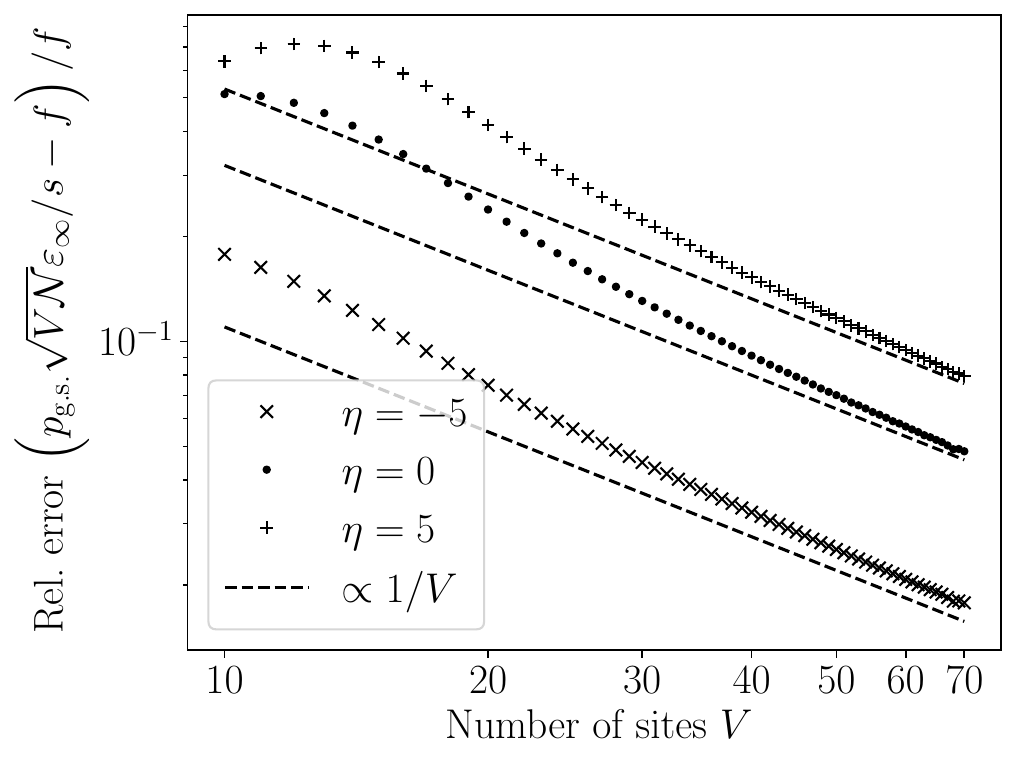}
    \caption{\textbf{Comparison of semianalytical $p_\gs$ for paramagnetic 1D Ising model to scaling form.}
    \textbf{Left:} error in scaling function \eqref{eq:scaling-form} compared to semianalytical $p_\gs$ as a function of of $\eta = \sqrt{V2^V} (E - E_c)/E_\infty$, across system sizes $V$.
    \textbf{Right:} error as a function of number of sites $V$ for three fixed $\eta$, one on each side of the transition. In both cases the error is shrinking like $1/V$, indicating that it is a finite size effect---likely from higher moments we neglect in deriving the scaling form.
    }
    \label{fig:rescaled-comparison}
\end{figure*}

Now compute $E_\av$ in terms of $p_\gs$ via $\beta$.
$E_\av$ is
\begin{align}
    E_\av &= \sum_{\alpha \ne gs} \frac {E_\alpha}{\lambda + \beta E_\alpha } \notag \\
    &= \sum_{\alpha \ne g.s.} \frac 1 {\lambda + \beta E_\alpha} + \sum_\alpha \frac {E_\alpha - 1}{\lambda + \beta + \beta (E_\alpha - 1)} \\
    &= (1 - p_\gs) + \frac {1 }{\lambda + \beta}\sum_\alpha (E_\alpha - 1) \sum_n \left[\frac{\beta(1-E_\alpha)}{\lambda + \beta}\right]^n  \notag
\end{align}
The sum over $n$ in the second term is a sum over central moments of the density of states. Since they drop off as $V^{-n/2}$, we keep only the leading correction, coming from the $n = 1$ term (second central moment) for
\begin{align}
    E_\av = (1 - p_\gs) - \frac {\mathcal N_\bulk \beta \mu_2} {(\lambda + \beta)^2} + O(\mu_3)
\end{align}
where $\mu_2, \mu_3$ are again the second and third central moments of $E_\alpha / E_\infty$ in the bulk density of states.
Using \eqref{eq:app-norm-pgs-beta} and \eqref{eq:pgs-beta} this becomes
\begin{align}
    p_\gs = &(1 - p_\gs)  \left(1 - \mu_2 \frac {p_\gs (\mathcal N_\bulk + 1) - 1}{\mathcal N_\bulk p_\gs} \right) \notag
\end{align}
Solve the resulting quadratic, dropping $O(s^4)$ and genuine $O(1/\mathcal N_\bulk)$ corrections---being careful to remember that $p_\gs$ may itself be $\sim 1/\mathcal N_\bulk$---for 
\begin{align}
\begin{split}
    p_\gs = \frac 1 2 \sqrt{\mu_2/\mathcal N_\bulk} \Big[-&(E-E_c) \sqrt{N/\mu_2}  \\
    &\qquad+ \sqrt{(E-E_c)^2 N/\mu_2 + 4}\Big]\notag
    \end{split}
\end{align}
with
\begin{align}
    E_c = 1 - \mu_2\;.
\end{align}
At this point we can return to units where $E_\infty \ne 1$.
Now
\begin{align}
    \mu_2 = s^2V\;,
\end{align}
where as usual $s^2 = V \mathcal N^{-1} \sum_\alpha (\varepsilon_\alpha - \varepsilon_\infty)^2 $
(and we neglect the small difference $\mathcal N_\bulk - \mathcal N$),
and
\begin{align}
\begin{split}
   E_c &= E_\infty\left(1 -  \frac{s^2V}{E_\infty^2} \right)\\
   \varepsilon_c &= \varepsilon_\infty \left( 1 - \frac {s^2}{V\varepsilon_\infty^2}\right)\;.
\end{split}
\end{align}
Writing 
\begin{align}
\begin{split}
    \eta
    &\equiv (E - E_c) \sqrt{\mathcal N/\mu_2} \\
    &= (E - E_c) \sqrt{\mathcal N/s^2V} \\
    &= (\varepsilon - \varepsilon_c) \sqrt{\mathcal NV/s^2}\;,
\end{split}
\end{align}
the ground state probability takes the form given in the main text:
\begin{align}\label{eq:scaling-form}
    p_\gs = \frac s {\varepsilon_\infty \sqrt{VN}} \frac 1 2 \Big[-\eta + \sqrt{\eta^2 + 4}\Big]
\end{align}
Fig.~\ref{fig:rescaled-comparison} compares this prediction to the semianalytical calculation of $p_\gs$ and $E_\av$ as functions of $\beta$;
Fig.~\ref{fig:rescaled-comparison} left shows the error as a function of $\eta$ across $V$, and Fig.~\ref{fig:rescaled-comparison} right shows the error at a few fixed $\eta$ as a function of volume. 
(In Fig.~\ref{fig:rescaled-comparison} we use a linear interpolation between the discrete $\beta$, hence $\varepsilon_\av$, at which we evaluate.
Note that in that plot we use the finite-size $E_c^{-1} = \frac 1 {\mathcal N_\bulk} \sum_{\alpha \ne \gs} E_\alpha^{-1}$ rather than the moment expansion value.)

\begin{figure*}
    \centering
    \includegraphics[width=0.75\textwidth]{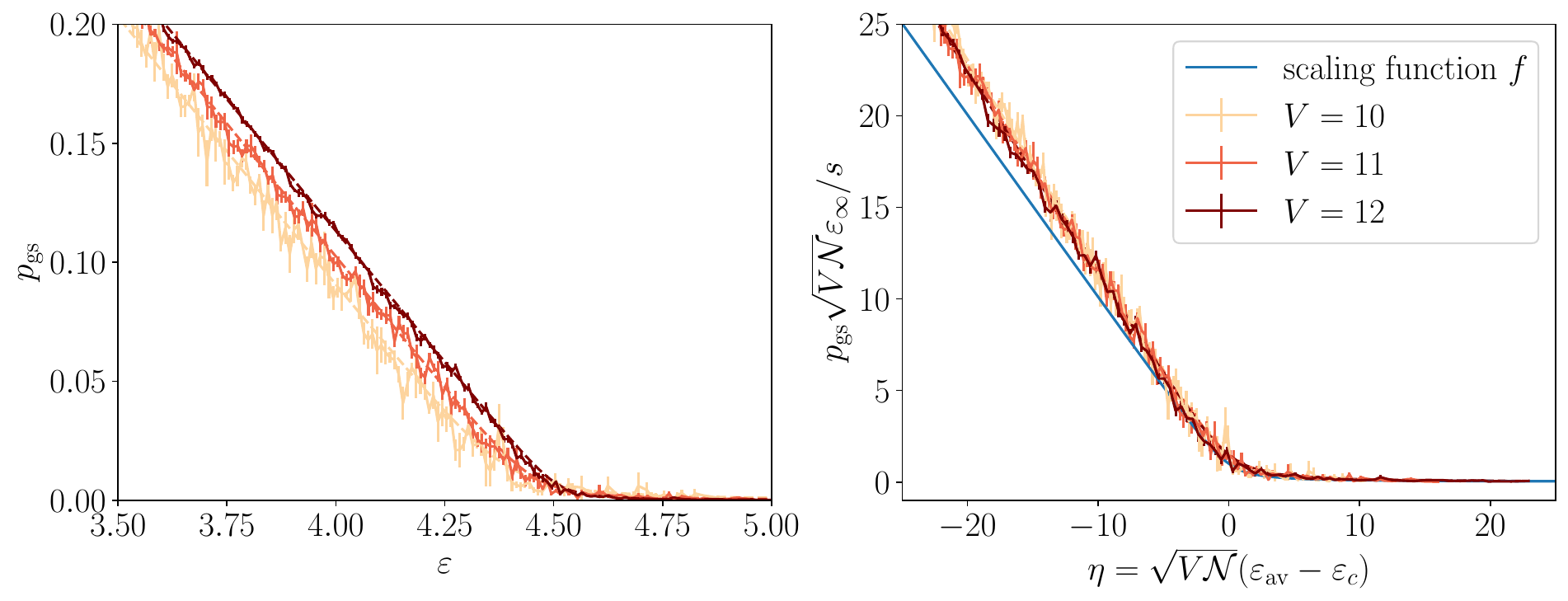}
    \caption{
    \textbf{Sampling data (solid / errorbars) and semianalytical free fermion} $p_\gs$, $E_\av$ for the 1D paramagnetic TFIM.
    Sampling data agrees with the semianalytical calculation up to sampling error and with the scaling form \eqref{eq:scaling-form} up to sampling error and finite size effects.
    }
    \label{fig:collapse_semianalytical_sampling}
\end{figure*}

Fig.~\ref{fig:collapse_semianalytical_sampling} compares the sampling data to the semianalytical $p_\gs(\beta), E_\av(\beta)$ computed as in App.~\ref{app:pgs-Eav-tfim}.
They agree to within sampling error.
The sampling data shows a good scaling collapse, though the apparent collapse function differs from the scaling function \eqref{eq:scaling-form} due to finite-size effects coming from higher central moments neglected in the derivation of \eqref{eq:scaling-form}.

\section{Gapless Hamiltonians}\label{app:gapless-hamiltonians}

We have argued that quantities like
\begin{align}
    2^{-V}\sum_\alpha \frac 1 E_\alpha \approx \int_0^{\|H\|_\infty} dE\; \rho(E) \frac 1 E
\end{align}
control the ground state condensation transition.
When the Hamiltonian is gapless the integral is not well defined (it is IR-divergent).
Does this change our results?
We expect not.
Gapless Hamiltonians generally have extensive high-temperature free energies,
so we expect the arguments of
App.~\ref{app:fe-mu} (in which we argued that high central moments of the energy density are small, justifying a moment expansion),
App.~\ref{app:Ec-system-size} (in which we used that moment expansion to estimate how $E_c$ depends on system size),
and
App.~\ref{app:scaling-collapse} (in which we used that moment expansion to predict a scaling collapse)
to hold.

Nonetheless it is enlightening to consider a concrete example.
Consider a system with low-energy single-excitation spectrum
\begin{align}
    E_k \sim \omega_D k^z\;,
\end{align}
where the ``Debye frequency'' $\omega_D$ sets the energy scale.
(We work in one dimension for simplicity, and assume periodic boundary conditions.)
Imagine the full Hamiltonian spectrum consists of $V$ low-energy single-excitation states,
together with $2^V-V$ many-excitation states with energy $E > \omega_D$, mean $E_\infty$, and variance $\sigma^2$.
Then
\begin{align}
\begin{split}
    &2^{-V} \sum_{\alpha \ne \gs} \frac 1 {E_\alpha} \approx 2^{-V} \sum_{k \ne 0} \frac 1 {\omega_D k^z} + \frac {1 - 2^{-V}V} {\sigma\sqrt{2\pi}} \int_{\omega_D}^\infty dE\; \frac 1 E e^{-(E - E_\infty)^2 / 2\sigma^2}\;.
\end{split}
\end{align}
(Assume $E_\infty - \omega_D \gg \sigma$, so restricting the Gaussian to $E \ge \omega_D$ does not appreciably change the normalization.)
The allowed $k$ are $k = 2\pi n / V$, $n = 1, \dots, V-1$, so the single-excitation contribution is
\begin{align}
\begin{split}
    2^{-V}\sum_{k \ne 0} \frac 1 {\omega_D k^z} &\propto 2^{-V} V^z / \omega_D \sum_{n = 1}^{V-1} \frac 1 {n^z} \\
    &\sim
    2^{-V}
    \begin{cases}
    V^z & z > 1 \\
    V \ln V & z = 1 \\
    V & z < 1
    \end{cases}\;.
\end{split}
\end{align}
In the large-system limit, then, this contribution from gapless single excitations is exponentially suppressed compared to the Gaussian contribution,
and indeed our results should be unchanged.
But for small systems it will contribute a subleading correction,
which may be appreciable.

We emphasize that this is at best a heuristic argument.
In particular, the real many-excitation spectrum will reach well below $\omega_D$
because it includes states with multiple long-wavelength, hence low-energy, excitations.

\end{document}